\documentclass[british,english,aps,prb,twocolumn,floatfix,letterpaper,superscriptaddress,amssymb,eprint]{revtex4-2}

\usepackage{graphicx}
\usepackage{dcolumn}
\usepackage{bm}
\usepackage{mathtools}
\usepackage{array}
\usepackage{blindtext}
\usepackage{comment}
\usepackage[T1]{fontenc}
\usepackage{braket}
\usepackage{dsfont}
\usepackage{svg}
\usepackage[normalem]{ulem}
\usepackage[unicode=true,pdfusetitle,
 bookmarks=true,bookmarksnumbered=false,bookmarksopen=false,
 breaklinks=false,pdfborder={0 0 1},backref=false,colorlinks=false]
 {hyperref}
 \hypersetup{colorlinks,allcolors=blue,breaklinks=true}
\allowdisplaybreaks[1]

\makeatother

\begin{document}

\newcommand{\Nin}{N_{\textrm{in}}}
\newcommand{\Nout}{N_{\textrm{out}}}
\newcommand{\p}{\prime}
\newcommand{\calT}{\mathcal{T}}
\newcommand{\calS}{\mathcal{S}}
\newcommand{\ii}{\mathrm{i}}
\newcommand{\id}{\mathrm{d}}
\newcommand{\A}[1]{\textcolor{red}{#1}}

\newcolumntype{P}[1]{>{\centering\arraybackslash}p{#1}}

\title{Photon-instanton scattering in a superconducting circuit: Beyond the very high impedance regime}

\author{Amir Burshtein}
\thanks{Equal contribution}
\email{burshtein2@mail.tau.ac.il}

\affiliation{Raymond and Beverly Sackler School of Physics and Astronomy, Tel Aviv
University, Tel Aviv 6997801, Israel}

\author{David Shuliutsky}
\thanks{Equal contribution}

\affiliation{Raymond and Beverly Sackler School of Physics and Astronomy, Tel Aviv
University, Tel Aviv 6997801, Israel}

\author{Roman Kuzmin}
\affiliation{Department of Physics, University of Wisconsin-Madison, Madison, WI 53706, USA}
\affiliation{Department of Physics, University of Maryland, College Park, MD 20742, USA}

\author{Vladimir E. Manucharyan}
\affiliation{\'Ecole Polytechnique F\'ed\'erale de Lausanne, 1015 Lausanne, Switzerland}
\affiliation{Department of Physics, University of Maryland, College Park, MD 20742, USA}

\author{Moshe Goldstein}
\affiliation{Raymond and Beverly Sackler School of Physics and Astronomy, Tel Aviv
University, Tel Aviv 6997801, Israel}

\begin{abstract}
Instantons, semi-classical trajectories of quantum tunneling in imaginary time, have long been used to study thermodynamic and transport properties in a myriad of condensed matter and high energy systems. A recent experiment in superconducting circuits [Phys. Rev. Lett. 126, 197701, (2021)] provided first evidence for direct dynamical signatures of instantons (phase slips), manifested by order-unity inelastic decay probabilities for photons with which they interact, motivating the development of a scattering theory of instantons [Phys. Rev. Lett. 126, 137701, (2021)]. While this framework successfully predicted the measured inelastic decay rates of the photons for several experimental devices, it is valid only if the tunneling time of the instantons is much shorter than the relaxation time due to the environment in which they are embedded, and requires a closed analytical expression for the instanton trajectory. Here, we alleviate these restrictions by incorporating numerical methods that eliminate some of the previously applied approximations. Our results improve the agreement with the experimental measurements, also for devices with lower impedances and thus shorter relaxation times, without fitting parameters. This framework should be useful in many other quantum field theoretical contexts.
\end{abstract}

\maketitle

\section{Introduction}

The path-integral formulation of quantum tunneling is based on the notion of instantons. These are solutions to the equations of motion in Euclidean spacetime, traversing a classically-forbidden path over a finite amount of time with a finite Euclidean action. Instantons are a quintessential non-perturbative tool in the analysis of both single-particle problems and quantum field theories, and are crucial for understanding paradigmatic condensed matter and high energy systems, such as the sine-Gordon \cite{zamolodchikov_factorized_1979} and spin-boson models \cite{leggett_dynamics_1987}, or the supersymmetric Yang-Mills theory \cite{gross_qcd_1981, vandoren_lectures_2008}. In the context of superconductivity and superfluidity, they are known as ``phase slips'' \cite{schon_quantum_1990, fazio_quantum_2001}, and have dramatic effects on phase diagrams \cite{schmid_diffusion_1983,bulgadaev_phase_1984, rastelli_quantum_2013, bard_superconductor-insulator_2017}. They have recently drawn interest also in the field of physical chemistry, where quantum tunneling allows for chemical reactions even at extremely low temperatures \cite{kastner_tunnelling_2020}.

\begin{figure}[b]
    \centering
    \includegraphics[width=0.98\columnwidth]{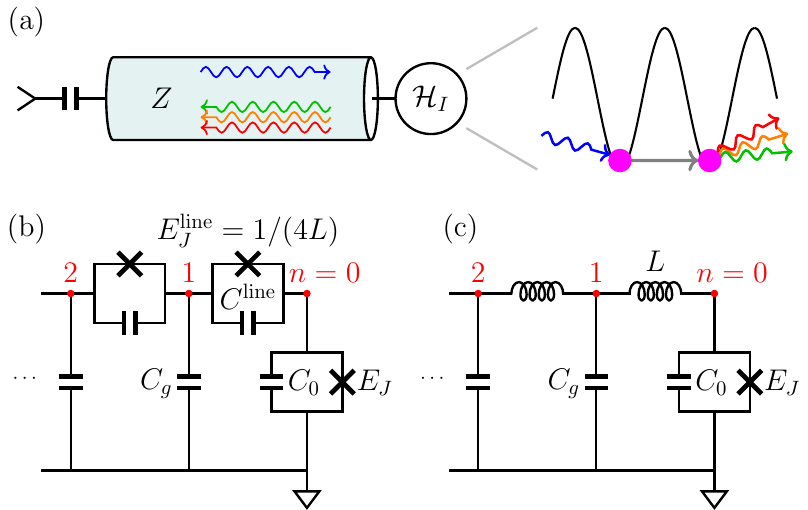}
    \caption{(a) A photon propagating in the transmission line triggers an instanton at the impurity $\mathcal{H}_I$, leading to the emission of multiple photons at lower frequencies. (b) Transmon ($n=0$) coupled to a Josephson array. (c) Transmon coupled to the effective linear transmission line with impedance $Z = \sqrt{L/C_g}$.}
    \label{fig:intro_fig}
\end{figure}

In accordance with naming conventions, instantons are often thought of as particle-like objects. However, since they are inherently defined in imaginary time, their observation in experiments through collisions with ordinary particles had not been considered possible, and the theoretical treatment of instantonic systems, commonly known as instanton calculus, mostly concerned thermodynamic and transport properties. Strikingly, a recent experiment in circuit quantum electrodynamics (cQED) showed that instantons may in fact imprint a dynamical signature on photons with which they interact. Depicted in Fig.~\ref{fig:intro_fig}, an array of Josephson junctions, serving as a waveguide, is galvanically-coupled to a transmon qubit \cite{koch_charge-insensitive_2007}, playing the role of an artificial atom with a periodic sinusoidal potential. This constitutes a cQED simulation of light-matter interaction with an effective fine-structure constant of order-unity, thanks to the high impedance of the Josephson array, of the order of the resistance quantum. The upshot is the possibility to gently probe the properties of the artificial atom by measuring the enhanced elastic and inelastic scattering rates of single-photons in the array, which was exploited in numerous recent experimental and theoretical works \cite{goldstein_inelastic_2013,kuzmin_superstrong_2019,puertas_martinez_tunable_2019, bard_decay_2018, wu_theory_2019, kuzmin_quantum_2019, houzet_critical_2020, burshtein_photon-instanton_2021, kuzmin_inelastic_2021, mehta_down-conversion_2023, mehta_theory_2022, kuzmin_observation_2023, burshtein_inelastic_2024, houzet_microwave_2024, remez_bloch_2024, burshtein_quantum_2024}. Indeed, measurements showed that a single-photon may split into lower-frequency photons with order-unity probability, far exceeding decay rates induced by other known photon-splitting mechanisms in nonlinear quantum optics \cite{guerreiro_nonlinear_2014} or nuclear physics \cite{akhmadaliev_experimental_2002}. The dominant source for this decay was found to be instantons occurring in the periodic potential of the transmon, in a process illustrated in Fig.~\ref{fig:intro_fig} --- a photon triggers a phase slip in the transmon, which in turn leads to the emission of multiple lower-frequency photons. Remarkably, the inelastic decay rate induced by the instantons exceeded other anharmonicities in the transmon by several orders of magnitude.

This scattering process highlighted a gap in instanton calculus, and motivated the development of a scattering theory of instantons. The main obstacle in developing such a theory was the analytical continuation of expressions obtained in imaginary time, where instantons are defined, to real time, where measurements are collected. Direct calculation of the real-time 2-point response function of the photons, which encodes the scattering rates, proved to be difficult (see Ref.~\cite{kolganov_real-time_2023} for a recent progress on the diagrammatic calculation of real-time response functions in the presence of instantons); a workaround was to calculate instead the $\calT$-matrix elements of the scattered photons, and in turn their inelastic decay rates by means of the Fermi golden rule \cite{burshtein_photon-instanton_2021}. The calculated decay rates agreed nicely with experimental measurements, without any fitting parameters \cite{kuzmin_inelastic_2021}, over a particular range of experimental parameters. While the formalism was developed in the scope of cQED, its applications could span many condensed matter and high energy systems.

The formalism devised in Ref.~\cite{burshtein_photon-instanton_2021} is analytical, and allows for the calculation of the inelastic decay rates of the photons without the need for numerical analytic continuation. However, resorting to purely analytical methods came at the cost of some approximations. Most notably, the set of Euclidean semi-classical equations of motion, determining the response of the photon field to the instanton occurring at the transmon, had to be linearized in order to obtain analytically-closed expressions in Matsubara frequency which were then conveniently analytically-continued to real frequencies. This linearization could be justified when the relaxation $RC$ (actually $ZC$) time defined by the transmon capacitance and the array impedance is substantially longer than the tunneling time of the instanton. However, it is no longer valid when the relaxation time is short enough, constituting a strong coupling between the photons and the instantons; indeed, the theoretical predictions deviated from the experimental measurements in those cases. Moreover, the method applies only for potential shapes where a closed analytical expression for an instanton is known.

In this work, we extend the applicability regime of the instanton scattering formalism by incorporating numerical methods replacing some of the approximations applied in Ref.~\cite{burshtein_photon-instanton_2021}. We solve the full set of Euclidean semi-classical equations numerically, in contrast to the analytical solution of the linearized equations obtained in Ref.~\cite{burshtein_photon-instanton_2021}. We then adjust our formalism to obtain the $\calT$-matrix elements, and subsequently the inelastic scattering rates of the photons, from this numerical solution. Importantly, we exploit the fact that the only nonlinearity in our system resides in the transmon (where the instanton occurs), and use our knowledge of the analytical structure of the approximate solution, to decrease the issue of potentially ill-posed numerical analytic continuation as much as possible. Our modified framework produces improved results with regards to the experimental data, and recovers the result of the original fully-analytical framework in the appropriate limit, where the relaxation time is long enough. Similarly to the fully-analytical framework, the applicability of the improved formalism is not restricted only to cQED systems, and it should be useful in many other quantum field theoretical contexts.

The rest of the paper is organized as follows. In Section \ref{sec:model}, we present the model considered in this work, which was realized experimentally in Ref.~\cite{kuzmin_inelastic_2021}. The numerical solution of the Euclidean semi-classical equations of motion of the array in response to an instanton at the transmon is discussed in Section \ref{sec:EOM}. Then, in Section \ref{sec:Tmat}, we show how to use this numerical solution to calculate the decay rates of the photons. While some steps are identical to those of Ref.~\cite{burshtein_photon-instanton_2021}, we present the full derivation for completeness, and emphasize the key differences between the numerical approach of this work and the analytical approach of Ref.~\cite{burshtein_photon-instanton_2021}. Section \ref{sec:results} shows the results of our method, illustrating enhanced decay rates at short relaxation $RC$ times, which improve the agreement with the experiment in regions not captured by the calculation of Ref.~\cite{burshtein_photon-instanton_2021}. We conclude in Section \ref{sec:conclusions}. Some technical details are relegated to the Appendices.

\section{Model}\label{sec:model}
The method we develop in this work is general, and not restricted to a specific model or platform. For concreteness, we focus on the system realized experimentally in Ref.~\cite{kuzmin_inelastic_2021}. Consider a one-dimensional array of $N+1 \gg 1$ superconducting grains with lattice spacing $a$, terminated by a transmon, depicted in Fig.~\ref{fig:intro_fig}. In the following, we work in units where $e=\hbar=a=1$; in particular, the superconducting flux quantum becomes $\Phi_0 = \hbar/(2e)=1/2$. The grains are connected via Josephson junctions with Josephson energy $E_J^\textrm{line}$ and capacitance $C^{\textrm{line}}$, and connected to the ground by capacitors $C_g$. The parameters of the Josephson junctions are chosen to suppress nonlinearities in the array, $E_J^\textrm{line}\gg E_C^\textrm{line}$ where $E_C^\textrm{line} = 1/(2 C^\textrm{line})$, such that the junctions act as effective inductors, with inductance $L=1/(4E_J^\textrm{line})$. The intergrain capacitance $C^{\textrm{line}}$ defines a plasma frequency, $\omega_p = 1/\sqrt{LC^{\textrm{line}}}$; in the following we consider energy scales significantly below $\omega_p$, and thus discard $C^{\textrm{line}}$ henceforth. Under these approximations, the Lagrangian reads
\begin{align}
\label{eq:LagSys}
\mathcal{L}=&\frac{C_0}{2}(\partial_t\phi_0)^2-E_J\left(1-\cos(2\phi_0)\right) \nonumber\\
&+\sum_{n=1}^N{\left[\frac{C_g}{2}(\partial_t\phi_n)^2-\frac{1}{2L}(\phi_n-\phi_{n-1})^2\right]},
\end{align}
where $\phi_n$ is the superconducting phase at the $n$th grain, and $E_J$ and $E_C = 1/(2C_0)$ are the Josephson and charging energies of the transmon, respectively. The quadratic one-dimensional transmission line is characterized by a velocity $v = 1/\sqrt{L C_g}$ and impedance $Z = \sqrt{L/C_g}$. In the following, we define the normalized impedance $z=Z/R_Q$, where $R_Q=h/(2e)^2=\pi/2$ is the superconducting resistance quantum. The kinetic inductance of the Josephson array enables the implementation of large impedances, $z \sim 1$, two orders of magnitude larger than impedances in typical transmission lines \cite{kuzmin_superstrong_2019, kuzmin_quantum_2019}. The order-unity normalized impedance is equivalent to an order-unity effective fine-structure constant, leading to strong light-matter interaction that allows for the high-probability splitting of single-photons \cite{kuzmin_inelastic_2021}.

The artificial atom in this cQED setup is the transmon, whose Josephson and charging energies are chosen such that $E_J \gtrsim E_C$. Expanding the cosine term of the transmon to harmonic order, $E_J(1-\cos(2\phi_0))\approx2E_J\phi_0^2$, gives rise to a resonance frequency, $\omega_0$. To lowest order in $E_C/E_J$, the resonance frequency of the transmon reads $\omega_0\approx\sqrt{8E_JE_C}$, and the first correction in perturbation theory is $\omega_0 \approx \sqrt{8 E_J E_C} - E_C$. An exact expression for the resonance frequency is given in terms of  the Mathieu characteristic values, as discussed in Appendix \ref{app:trans}. Let us note that the resonance frequency may be measured directly in the experiment \cite{kuzmin_superstrong_2019}; therefore, in the following, we refer explicitly to $\omega_0$. Also note that the transmon is realized by a SQUID, such that $E_J$ and $\omega_0$ can be tuned by an external magnetic field. 

The harmonic expansion of the transmon leads to a quadratic Lagrangian which can be diagonalized by plane waves (i.e. photons), $\phi_n=\sum_k{\sin(nk+\delta_k)\phi_k}$. Here $\phi_k$ are the system eigenstates satisfying $\partial_t^2\phi_k=-\omega_k^2\phi_k$, with an approximately linear dispersion relation, $\omega_k=2v\sin(k/2)\approx vk$ (the array velocity $v$ is assumed to be much larger than any other energy scale). The phase shift reads
\begin{equation}
\label{eq:delta_k}
    \tan{\delta_k}=\frac{\omega_k\Gamma_0\sqrt{1-\left(\frac{\omega_k}{2v}\right)^2}}{\omega_0^2-\left(1-\frac{\Gamma_0}{2v}\right)\omega_k^2}\approx\frac{\Gamma_0\omega_k}{\omega_0^2-\omega_k^2},
\end{equation}
where $\Gamma_0=1/(ZC_0)$ is the inverse $RC$ time defined by the array impedance and the transmon capacitance, which sets the elastic broadening of the transmon frequency $\omega_0$. The finite array size quantizes the $k$ modes, $k(N+1/2) + \delta_k = \pi(m + 1/2)$ with $m \in \mathbb{N}$, giving rise to a mode spacing $\Delta = \pi v/N$. The diagonalized Lagrangian reads
\begin{equation}
    \mathcal{L} \approx \sum_k \left[\frac{C_k}{2}\left(\partial_t \phi_k\right)^2 - \frac{C_k\omega_k^2}{2}\phi_k^2\right],
\end{equation}
where $C_k \approx N C_g/2$ is the capacitance of the $k$ mode.

In the following, it would be useful to consider another basis of modes, which diagonalizes the bulk part ($n \ge 1$) of the Lagrangian in Eq.~\eqref{eq:LagSys}, that is, without the transmon. Note that the bulk is quadratic even without the harmonic approximation of the transmon. To distinguish these modes from the $\phi_k$ modes of the full linearized system, we decorate the bulk modes with a hat and denote them by $q$, $\phi_n=\sum_q{\sin(q(n-1)+\hat{\delta}_q)\hat{\phi}_q}$ for $n \ge 1$, where $\partial_t^2\hat{\phi}_q=-\omega_q^2\hat{\phi}_q$, and
\begin{equation}
    \label{eq:delta_q}
    \tan{\hat{\delta}_q}=\frac{\omega_qv\sqrt{1-\left(\frac{\omega_q}{2v}\right)^2}}{v^2-\omega_q^2/2}.
\end{equation}
The dispersion relation is the same as that of the full modes, $\omega_q = 2v\sin(q/2)$, but the phase shift $\hat{\delta}_q$ differs from $\delta_k$, giving rise to a different quantization relation, $q(N-1/2) + \hat{\delta}_q = \pi(m + 1/2)$. Using the bulk modes, the Lagrangian may be written as
\begin{align}
    \label{eq:LagBulk}
\mathcal{L}=&\frac{C_0}{2}(\partial_t\phi_0)^2-E_J\left(1 - \cos(2\phi_0)\right)-\frac{1}{2L}\phi_0^2 \nonumber \\
&\hspace{-1cm}+\sum_q{\left[\frac{C_q}{2}(\partial_t\hat{\phi}_q)^2-\frac{C_q\omega_q^2}{2}\hat{\phi}_q^2+\frac{1}{L}\phi_0\hat{\phi}_q\sin{\hat{\delta}_q}\right]},
\end{align}
where $C_q \approx N C_g/2$ as well. Let us stress that the modes measured by spectroscopy are those of the full system, $\phi_k$ \cite{kuzmin_superstrong_2019}; however, the bulk modes $\hat{\phi}_q$ would be very useful in the following sections.

\section{Instantons in the coupled transmon-array system}\label{sec:EOM}

Consider the Lagrangian in Eq.~\eqref{eq:LagSys}. Applying a Wick rotation, $\tau = \ii t$, the Euler-Lagrange equations read
\begin{align}
    \label{eq:full_eq_system}
    &\partial_\tau^2\phi_0=\frac{\omega_0^2}{2}\sin(2\phi_0)-v\Gamma_0(\phi_1-\phi_0),\nonumber\\
    &\partial_\tau^2\phi_n=v^2(2\phi_n-\phi_{n-1}-\phi_{n+1}), &(1\leq n < N)\nonumber\\
    &\partial_\tau^2\phi_N=v^2(\phi_N-\phi_{N-1}).
\end{align}
An instanton solution at $n=0$ corresponds to a single tunneling event between two adjacent minima in the sinusoidal potential, such that $\phi_0(\tau\rightarrow -\infty) = 0$ and $\phi_0(\tau\rightarrow\infty) = \pm \pi$, where the plus and minus signs correspond to an instanton and an anti-instanton, respectively. In the absence of an environment, the Euler-Lagrange equation for a decoupled transmon reads $\partial_\tau^2\phi_0 = \left(\omega_0^2/2\right) \sin(2\phi_0)$, and the well-known analytical instanton solution is \cite{rajaraman_solitons_1982}
\begin{equation}\label{eq:phi0-analytical}
    \pm\phi_0^{(0)}(\tau) = \pm2\arctan(\exp(\omega_0\tau)),
\end{equation}
tunneling across a characteristic time scale $1/\omega_0$,  with the Fourier transform $\phi_0^{(0)}(\ii\omega)=\pi/(\ii\omega\cosh(\pi\omega/(2\omega_0)))$. When the instanton is coupled to the environment, the phases $\phi_{n\ge 1}$ must have the same values at $\tau \rightarrow \pm \infty$ as $\phi_0$, in order to minimize the Euclidean action. Our goal is find the joint semi-classical trajectories of the instanton at $n=0$ and the response of the array. In the following, we seek a solution for the coupled equations in Eq.~\eqref{eq:full_eq_system}, where all phases satisfy $\phi_n(\tau\rightarrow -\infty)=0$ and $\phi_n(\tau\rightarrow \infty)=\pi$.

In order to avoid a computationally-costly solution of the full set of $N + 1$ coupled second-order differential equations, we exploit the fact that the Lagrangian is nonlinear only at the transmon. From the Wick-rotated version of Eq.~\eqref{eq:LagBulk}, we find
\begin{equation}
    \label{eq:bulk-phi0}
\hat{\phi}_q(\ii\omega)=\frac{v}{ZC_q}\frac{\sin{\hat{\delta}_q}}{\omega^2+\omega_q^2}\phi_0(\ii\omega),
\end{equation}
which shows that the solution for $\phi_0$ determines the bulk modes $\hat{\phi}_q$. In particular, we can find the solution at $n=1$,
\begin{equation}
    \label{eq:phi_1-phi_0}
\phi_1(\ii\omega)=\sum_q{\frac{v}{ZC_q}\frac{\sin^2{\hat{\delta}_q}}{\omega^2+\omega_q^2}\phi_0(\ii\omega)}\approx\left(1-\frac{|\omega|}{v}\right)\phi_0(\ii\omega),
\end{equation}
where the sum over $q$ was evaluated as an integral in the thermodynamic limit, $\sum_q\rightarrow (N/\pi) \int_0^{\pi} \id q$. Eq.~\eqref{eq:full_eq_system} can now be solved iteratively. We initialize the solution for $\phi_0$ with that of an isolated transmon, $\phi_0^{(0)}$, given by Eq.~\eqref{eq:phi0-analytical}. At each step $m\ge 1$, we use the result of the previous step $\phi_0^{(m-1)}$ in Eq.~\eqref{eq:phi_1-phi_0} to find $\phi_1^{(m)}(\ii\omega)=(1-|\omega|/v)\phi_0^{(m-1)}(\ii\omega)$. Then, $\phi_1^{(m)}(\tau)$ (transformed back from Matsubara frequency to imaginary time) acts as an external force in the first line of Eq.~\eqref{eq:full_eq_system}, which may be solved numerically as an ordinary differential equation to yield $\phi_0^{(m)}$ with the boundary conditions $\phi_0^{(m)}(\tau \rightarrow -\infty)=0$ and $\phi_0^{(m)}(\tau \rightarrow \infty)=\pi$. We repeat this process several times until convergence.

Alternatively, it is possible to derive a single integro-differential equation for $\phi_0$. It is shown in Appendix \ref{app:kernel} that
\begin{equation}
    \label{eq:integro-diff-kernel}
        \partial_{\tau}^2\phi_0(\tau)=\frac{\omega_0^2}{2}\sin(2\phi_0(\tau))+v\Gamma_0\phi_0(\tau)-K(\tau)*\phi_0(\tau),
\end{equation}
where $*$ denotes convolution, and the kernel $K(\tau)$ is 
\begin{align}
\label{eq:kernel_full}
    K(\tau)=&v\Gamma_0\left(\frac{L_2(2v|\tau|)-I_2(2v|\tau|)}{|\tau|}+\frac{4v}{3\pi}\right),
\end{align}
where $L_2(x)$ is the modified Struve function and $I_2(x)$ is the modified Bessel function of the first kind \cite{abramowitz_handbook_1965}. For $v|\tau|\gg1$ we have $K(|\tau|\gg1/v)\approx\Gamma_0/(\pi \tau^2)$, which is the well-known friction kernel \cite{leggett_dynamics_1987, feynman_theory_1963}. Eq.~\eqref{eq:integro-diff-kernel} can be solved with appropriate methods developed in that context \cite{gelmi_idsolver_2014}.

\begin{figure}
    \centering
    \includegraphics[width=0.98\columnwidth]{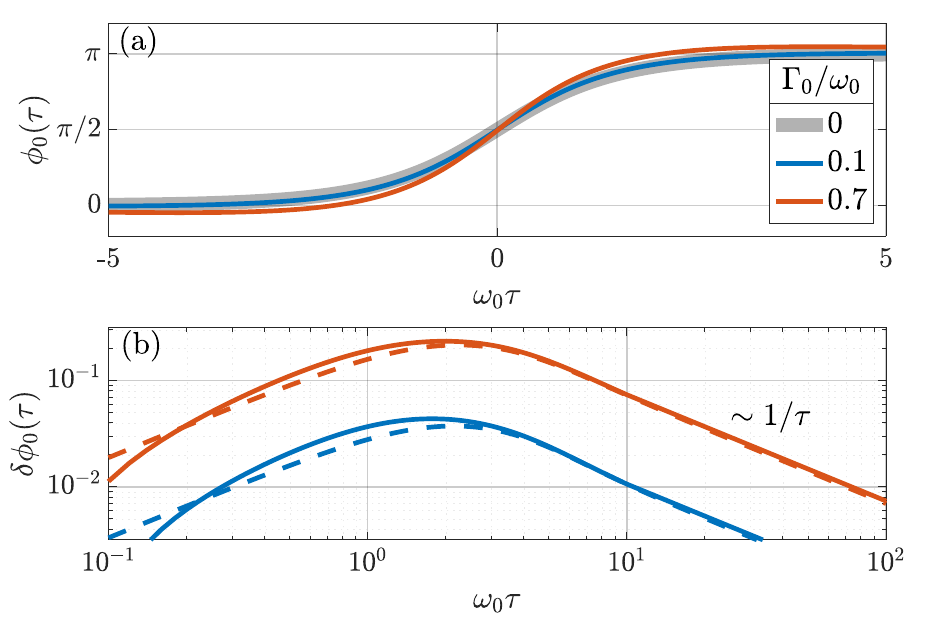}
    \caption{(a) The instanton trajectory $\phi_0(\tau)$ for an isolated transmon (thick gray line) and a transmon connected to a transmission line with two values of $\Gamma_0/\omega_0$. (b) The deviation from the free instanton trajectory, $\delta\phi_0(\tau) = \phi_0(\tau)-\phi_0^{(0)}(\tau)$, for the full (continuous lines) and approximated (dashed lines) equations, in log-log scale for $\tau > 0$ (the deviation is antisymmetric, $\delta\phi_0(-\tau)=-\delta\phi_0(\tau)$).}
    \label{fig:phi0_time}
\end{figure}

The deviation of the instanton trajectory at $n = 0$ from the bare trajectory, $\delta\phi_0(\tau) = \phi_0(\tau) - \phi_0^{(0)}(\tau)$, is displayed in Fig.~\ref{fig:phi0_time}. The result is also compared to the approximate analytical solution of Eq.~\cite{burshtein_photon-instanton_2021}, discussed in Appendix \ref{app:PRL_lim}. At large imaginary times, $v|\tau| \gg 1$, one finds the asymptotic behavior $\delta\phi_0 \sim 1/\tau$, which can be inferred from the approximate analytical solution of Ref.~\cite{burshtein_photon-instanton_2021}, as discussed in Appendix \ref{app:phi0}. Appendix \ref{app:phi0} also shows how to treat this slow asymptotic decay with care to avoid issues with the numerical analytic continuation that is discussed in Section \ref{sec:Tmat}.

\section{Obtaining the $\mathcal{T}$-matrix elements and the decay rate}\label{sec:Tmat}

The experiment in Ref.~\cite{kuzmin_inelastic_2021} showed that instantons are the dominant source of inelastic photon decay (i.e. the $\phi_k$ modes). We proceed to obtain the $\mathcal{T}$-matrix elements, required for the calculation of the inelastic decay rates of the photons using the Fermi golden rule. The $\mathcal{T}$-matrix elements quantify the transition amplitudes of photon scattering events in real time; to obtain them, one must perform analytical continuation for the solutions in imaginary time obtained in Section \ref{sec:EOM}. This procedure is very similar to the one applied in Ref.~\cite{burshtein_photon-instanton_2021}, with one crucial difference --- since the solutions for the set of semi-classical Euclidean equations were obtained numerically, it is necessary to apply numerical analytic continuation, a notoriously ill-posed problem \cite{silver_maximum-entropy_1990, jarrell_bayesian_1996}. Here, we show how knowledge of the analytical structure of the approximate solution, obtained in Ref.~\cite{burshtein_photon-instanton_2021}, is paramount in avoiding the obstacles associated with numerical analytic continuation.

In the following, we retrace the derivation of Ref.~\cite{burshtein_photon-instanton_2021} and point out the differences when $\phi_0(\tau)$ is obtained numerically rather than analytically. Some technical details are relegated to the Appendices.

Our goal is to obtain the $\calT$-matrix elements between $\Nin$ incoming photons with momenta $k_1^\p,\ldots,k_{\Nin}^\p$ and $\Nout$ outgoing photons with momenta $k_1,\ldots,k_{\Nout}$. This is possible by means of the Lehmann-Symanzik-Zimmermann (LSZ) reduction formula \cite{peskin_introduction_2018}, which relates the $\calS$-matrix elements, $\calS_{k_1,\ldots,k_{\Nout}}^{k_1^\p,\ldots,k_{\Nin}^\p}=\braket{k_1,\ldots,k_{\Nout}|k_1^\p,\ldots,k_{\Nin}^\p}$, to an imaginary time-ordered multipoint correlator of the full array modes $\phi_k$ with its external single-particle legs amputated. We focus on the single-instanton sector, neglecting contributions from scattering processes involving two or more instantons:
\begin{widetext}
\begin{equation}\label{eq:Smat}
    \calS_{k_1,\ldots,k_{\Nout}}^{k_1^\p,\ldots,k_{\Nin}^\p} = \lim_{\substack{\{\omega_j^\p\rightarrow\ii\omega_{k_j^\p}\}\\{\{\omega_j\rightarrow -\ii\omega_{k_j}\}}}}
    \prod_{j=1}^{\Nin} \frac{C_{k_j^\p}\left(\omega_j^{\p 2} + \omega_{k_j^\p}^2\right)}{\sqrt{2C_{k_j^\p} \omega_{k_j^\p}}}
    \prod_{j=1}^{\Nout} \frac{C_{k_j}\left(\omega_j^2 + \omega_{k_j}^2\right)}{\sqrt{2C_{k_j} \omega_{k_j}}}
    \left\langle\prod_{j=1}^{\Nin} \phi_{k_j^\p} (\ii\omega_j^\p)\prod_{j=1}^{\Nout}\phi_{k_j} (\ii\omega_j)\right\rangle_{\textrm{1-instanton}}.
\end{equation}
The multipoint correlator is calculated within the saddle-point approximation. The instanton contributes an amplitude $\lambda_0/2\times e^{-\delta S}$. As discussed in Appendix \ref{app:trans}, $\lambda_0$ is the charge dispersion of the lowest Bloch band of the transmon, stemming from the classical action $S_0$ of the isolated instanton and the Gaussian fluctuations around it. $\delta S$, to be derived below, corresponds to the response of the array to the instanton. To leading order in $E_C/E_J$ and $\Gamma_0/\omega_0$, we may replace the fields $\phi_k$ by their values at the classical instanton path, such that
\begin{equation}\label{eq:inst_corr}
    \left\langle\ldots\right\rangle = (\pm 1)^{\Nin+\Nout} \phi_{k_1^\p}(\ii\omega_1^\p)\ldots\phi_{k_{\Nin}^\p}(\ii\omega_{\Nin}^\p) \phi_{k_1}(\ii\omega_1)\ldots\phi_{k_{\Nout}}(\ii\omega_{\Nout}) \frac{\lambda_0}{2}e^{-\delta S},
\end{equation}
where the plus and minus signs correspond to instantons and anti-instantons, respectively. We then use the relation between the $\calS$-matrix and the $\calT$-matrix,
\begin{equation}
    \calS_{k_1,\ldots,k_{\Nout}}^{k_1^\p,\ldots,k_{\Nin}^\p} = \mathds{1} - 2\pi\ii\delta\left(\omega_{k_1}+\ldots+\omega_{k_{\Nout}} - \omega_{k_1^\p} - \ldots - \omega_{k_{\Nin}^\p}\right) \calT_{k_1,\ldots,k_{\Nout}}^{k_1^\p,\ldots,k_{\Nin}^\p}.
\end{equation}
Note that the identity part of the $\calS$-matrix is absent from the LSZ formula, and must be added by hand. For on-shell processes, $\omega_{k_1}+\ldots+\omega_{k_{\Nout}} = \omega_{k_1^\p} + \ldots + \omega_{k_{\Nin}^\p}$, the delta function is replaced by the single-mode density $1/\Delta$. Using Eq.~\eqref{eq:inst_corr} and taking the limits $\omega_j\rightarrow -\ii\omega_{k_j},\omega_j^\p\rightarrow\ii\omega_{k_j^\p}$, we find (up to unimportant global phase factors)
\begin{equation}
    \label{eq:Tmat2}
    \calT_{k_1,\ldots,k_{\Nout}}^{k_1^\p,\ldots,k_{\Nin}^\p} = (\mp 1)^{\Nin} (\pm 1)^{\Nout} f_{k_1^\p}\ldots f_{k_{\Nin}^\p} f_{k_1} \ldots f_{k_{\Nout}} \frac{\lambda_0}{2} e^{-\delta S},
\end{equation}
\end{widetext}
where
\begin{equation}
    \label{eq:fkDef}
    f_k = \lim_{\omega\rightarrow -\ii\omega_k} \sqrt{\frac{C_k}{2\omega_k}} \left(\omega^2 + \omega_k^2\right) \phi_k(\ii\omega).
\end{equation}
We consider first the correction to the action, $\delta S$, stemming from the response of the array to the instanton at $n=0$, and then evaluate the $f_k$ factors, quantifying the overlap of the instanton with the $k$ modes of the full system.

\subsection{Correction to the action $\delta S$}\label{subsec:dS}
Consider the Lagrangian in Eq.~\eqref{eq:LagBulk}, transformed to imaginary time. Using Eq.~\eqref{eq:bulk-phi0} to express $\hat{\phi}_q(\ii\omega)$ in terms of $\phi_0(\ii\omega)$, we find the Euclidean action
\begin{align}
    S =& \int_{-\infty}^{\infty }\id\tau \left[\frac{C_0}{2}(\partial_\tau\phi_0(\tau))^2 - E_J\cos\left(2\phi_0(\tau)\right)\right] \nonumber \\
    &+\frac{1}{2L}\int_{-\infty}^{\infty}\frac{\id\omega}{2\pi}\left[1 - \sum_q \frac{v \sin^2\hat{\delta}_q}{Z C_q\left(\omega^2+\omega_q^2\right)}\right]\left|\phi_0(\ii\omega)\right|^2.
\end{align}
The sum over $q$ was evaluated in Eq.~\eqref{eq:phi_1-phi_0}. It is useful to write $\phi_0 = \phi_0^{(0)}+\delta\phi_0$, which allows us to separate the action of the isolated instanton $\phi_0^{(0)}$ from the response of the array. We arrive at $S=S_0 + \delta S_1 + \delta S_2$, where $S_0=\sqrt{8 E_J/E_C}$  is the action of an isolated instanton, and 

\begin{align}\label{eq:dS}
    \delta S_1 =& \frac{C_0}{2}\int_{-\infty}^{\infty}\frac{\id\omega}{2\pi}\left[\left(\omega^2 + \omega_0^2\right)\left|\delta\phi_0(\ii\omega)\right|^2 + \Gamma_0\left|\omega\right|\left|\phi_0(\ii\omega)\right|^2\right], \nonumber \\
    \delta S_2 =&-E_J\int_{-\infty}^{\infty}\id\tau\Big[\cos\left(2\phi_0(\tau)\right)-\cos\left(2\phi_0^{(0)}(\tau)\right) \nonumber \\
    &\hspace{1cm}+ 2\sin\left(2\phi_0^{(0)}(\tau)\right)\delta\phi_0(\tau) + 2\delta\phi_0^2(\tau)\Big].
\end{align}
To make contact with the expressions obtained in Ref.~\cite{burshtein_photon-instanton_2021}, we rewrite the integral in $\delta S_1$ as a sum over the modes of the full system (in the thermodynamic limit). We find $\delta S_1 = \sum_k \tilde{f}_k^2/2$, with
\begin{equation}\label{eq:fkTilde}
    \tilde{f}_k = \tilde{f}_k^{\textrm{apprx}} \times \sqrt{\frac{\omega_k^2 + \omega_0^2}{\Gamma_0 \omega_k}\left|\frac{\delta\phi_0(\ii\omega_k)}{\phi_0^{(0)}(\ii\omega_k)}\right|^2+\left|1+\frac{\delta\phi_0(\ii\omega_k)}{\phi_0^{(0)}(\ii\omega_k)}\right|^2},
\end{equation}
where $\tilde{f}_k^{\textrm{apprx}} = \sqrt{2\Delta/(z\omega_k)}\times 1/\cosh(\pi\omega_k/(2\omega_0))$ is the corresponding expression obtained in Ref.~\cite{burshtein_photon-instanton_2021} using the linearized equations of motion. $\delta S_2$ is an additional correction to the action that vanishes in Ref.~\cite{burshtein_photon-instanton_2021}. We verify in Appendix \ref{app:PRL_lim} that the correction to the action converges to the result of Ref.~\cite{burshtein_photon-instanton_2021} when the appropriate limit is applied; that is, $\tilde{f}_k\rightarrow\tilde{f}^{\textrm{apprx}}_k$, and $\delta S_2 \rightarrow 0$.

The back-action of the array modes in response to the instanton leads to a renormalization of the charge dispersion of the lowest Bloch band, $\lambda_0 \rightarrow e^{-\delta S} \lambda_0$. Instantons are relevant for z > 1, giving rise to a characteristic scale $\lambda^{\star}\sim(\lambda_0/\omega_0^{1/z})^{1/(1 - 1/z)}$ below which instanton effects are nonperturbative (see Appendix \ref{app:trans} for a short discussion). In the following, we avoid this regime by restricting the discussion to higher energies, $\max\left\{\omega_k,T,\Gamma_0\right\} \gg \lambda^{\star}$.

\subsection{The $f_k$ factors}\label{subsec:fk}
Consider Eq.~\eqref{eq:fkDef}. Inverting $\phi_n = \sum_k \sin(kn + \delta_k)\phi_k$, we have
\begin{equation}
    \phi_k = \frac{C_0}{C_k} \sin\delta_k \phi_0 + \frac{C_g}{C_k} \sum_n \sin(kn + \delta_k) \phi_n.
\end{equation}
Writing $\phi_n = \sum_q \sin(q(n-1)+\hat{\delta}_q)\hat{\phi}_q$ for $n\ge 1$ and using Eq.~\eqref{eq:bulk-phi0}, we find
\begin{align} \label{eq:fkDefSum}
    f_k =& \lim_{\omega\rightarrow -\ii\omega_k} \frac{\omega^2 + \omega_k^2}{\sqrt{2C_k\omega_k}} \phi_0(\ii\omega) \times \sum_{n=1}^N \sin(kn+\delta_k) \nonumber \\
    &\times \frac{2}{LN}\sum_q \frac{\sin\hat{\delta}_q}{\omega^2+\omega_q^2} \sin\left(q(n - 1)+\hat{\delta}_q\right).
\end{align}
Note that the $n=0$ term is absent from the sum, since it vanishes in the limit $\omega\rightarrow -\ii\omega_k$. In order to proceed, one needs to evaluate the double summation over $n$ and $q$. While direct evaluation is possible, it is cumbersome, since it involves both the $k$ modes of the full system, as well as the $q$ modes of the array without the transmon. Fortunately, a convenient shortcut is provided by the result obtained for the approximate analytical solution in Ref.~\cite{burshtein_photon-instanton_2021}. Note that Eq.~\eqref{eq:fkDefSum} applies regardless of the manner in which $\phi_0(\ii\omega)$ is obtained. In Appendix \ref{app:PRL_lim}, we show that within the approximations of Ref.~\cite{burshtein_photon-instanton_2021}, we have
\begin{equation}\label{eq:phik_apprx}
    \phi_k^{\textrm{apprx}}(\ii\omega) = \frac{\omega_k \cos\delta_k}{Z C_k} \frac{1}{\omega^2 +\omega_k^2} \phi_0^{(0)}(\ii\omega),
\end{equation}
leading to (via Eq.~\eqref{eq:fkDef})
\begin{equation}
    f_k^{\textrm{apprx}} =\lim_{\omega\rightarrow -\ii\omega_k} \frac{1}{\pi}\sqrt{\frac{2\Delta\omega_k}{z}}\cos\delta_k\phi_0^{(0)}(\ii\omega).
\end{equation}
On the other hand, within the same approximations, it is shown in Appendix \ref{app:PRL_lim} that
\begin{equation} \label{eq:phi0_apprx}
    \phi_0^{\textrm{apprx}}(\ii\omega) = \frac{\omega^2+\omega_0^2}{\omega^2+\Gamma_0|\omega|+\omega_0^2} \phi_0^{(0)}(\ii\omega).
\end{equation}
Comparing with Eq.~\eqref{eq:fkDefSum}, we immediately find that
\begin{align}
    \label{eq:fk}
    f_k =&\frac{1}{\pi}\sqrt{\frac{2\Delta\omega_k}{z}}\cos\delta_k\lim_{\omega\rightarrow -\ii\omega_k} \left[\frac{\omega^2 +\Gamma_0 |\omega| +\omega_0^2}{\omega^2 + \omega_0^2} \phi_0(\ii\omega)\right] \nonumber \\
    =&f_k^{\textrm{apprx}} \times \lim_{\omega \rightarrow -\ii\omega_k} \left[\frac{\omega^2+\Gamma_0|\omega|+\omega_0^2}{\omega^2+\omega_0^2}  \frac{\phi_0(\ii\omega)}{\phi^{(0)}_0(\ii\omega)}\right],
\end{align}
where $f_k^{\textrm{apprx}}$ is the factor obtained in Ref.~\cite{burshtein_photon-instanton_2021},
\begin{equation} \label{eq:fk_apprx}
    f_k^{\textrm{apprx}} = \sqrt{\frac{2\Delta}{z \omega_k}} \frac{\omega_0^2 - \omega_k^2}{\cos\left(\frac{\pi\omega_k}{2\omega_0}\right)\sqrt{\left(\omega_0^2-\omega_k^2\right)^2 + \left(\Gamma_0\omega_k\right)^2}}.
\end{equation}
Note that the first line of Eq.~\eqref{eq:fk} holds for any potential shape at $n=0$, where $\omega_0$ is the resonance frequency of the expanded harmonic potential around the minima and $\Gamma_0$ is the inverse $RC$ time. In the second line of Eq.~\eqref{eq:fk} we use the explicit form of $\phi_0^{(0)}(\ii\omega)$ for a cosine potential to relate the $f_k$ factors to $f_k^{\textrm{apprx}}$ obtained in Ref.~\cite{burshtein_photon-instanton_2021}.

This second line of Eq.~\eqref{eq:fk} is favorable for two reasons. First, only a single analytical continuation of the numerical data is required to obtain the $f_k$ factors for all $k$ modes. Second, it is reasonable to assume that $\phi_0(\ii\omega)$ should not deviate too significantly from the approximate solution in Eq \eqref{eq:phi0_apprx}, such that its analytical structure, namely, the number of zeros and poles and their position in the complex plane, should be similar to that of the full numerical solution. This is crucial to avoid problems that could emerge from numerical inaccuracies in the estimation of the poles of $\phi_0(z)$ in the complex plane. Importantly, note the $\cos\delta_k$ factor in the first line of Eq.~\eqref{eq:fk}, that approaches zero as $\omega_k\rightarrow\omega_0$, and counters the pole at $\ii\omega = \omega_0$ of $\phi_0^{(0)}(\ii\omega)$ to give a finite on-resonance value. A numerical error in the evaluation of the pole of the term inside the brackets in the first line of Eq.~\eqref{eq:fk} would lead to divergences in $f_k$ near resonance. The second line of Eq.~\eqref{eq:fk} removes this threat, by preemptively taking care of the pole at $\ii\omega = \omega_0$.

\begin{figure}
	\centering
	\includegraphics[width=0.98\columnwidth]{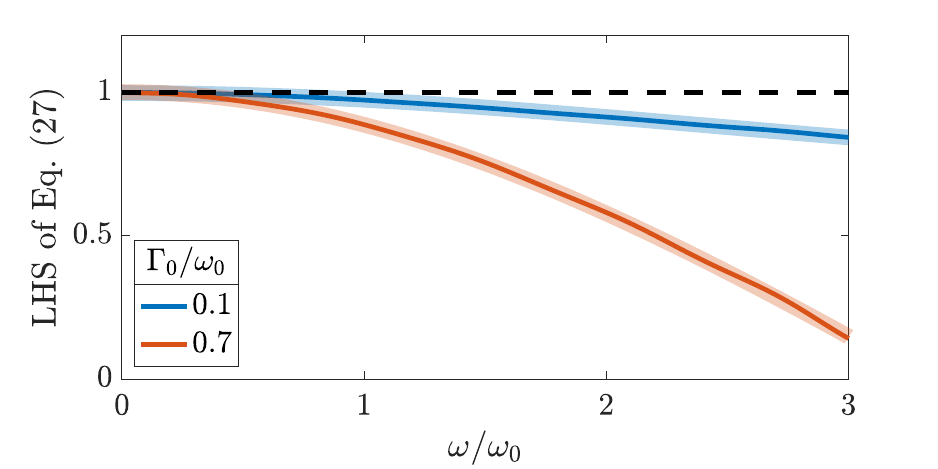}
	\caption{The argument on the left-hand side of Eq.~\eqref{eq:phi0_cont} (thin continuous lines) and the fit to a polynomial in Matsubara frequencies (thick transparent lines) using $p=6$. The dashed black line at unity corresponds to the left-hand side of Eq.~\eqref{eq:phi0_cont} for the approximate solution.}
	\label{fig:phi0_freq}
\end{figure}

The anticipated analytical structure of the solution $\phi_0(\ii\omega)$ allows us to use a very simple method for numerical analytic continuation. The combination inside the brackets in Eq.~\eqref{eq:fk} is not expected to have any poles. Therefore, in order to obtain $f_k$, we fit the expression inside the brackets in Eq.~\eqref{eq:fk} to a polynomial in Matsubara frequency, rather than a Pad\'e approximant. Namely, we write
\begin{equation}\label{eq:phi0_cont}
    \frac{\omega^2+\Gamma_0|\omega|+\omega_0^2}{\omega^2+\omega_0^2} \times \frac{\phi_0(\ii\omega)}{\phi^{(0)}_0(\ii\omega)} = \sum_{l=0}^p \alpha_l \omega^l,
\end{equation}
where $p$ is the order of the fitting polynomial, and estimate the coefficients $\alpha_l$ from the numerical data. Note that since $\phi_0(\tau)$ is real, $\phi_0(\ii\omega)$ is an even function of $\omega$, such that $\alpha_{1,3,\ldots}=0$. Fig.~\ref{fig:phi0_freq} shows an example of the left-hand side of Eq.~\eqref{eq:phi0_cont} and the fit on the right-hand side. For $\Gamma_0 \ll \omega_0$, the left-hand side of Eq.~\eqref{eq:phi0_cont} approaches unity, the expected result for the approximation employed in Ref.~\cite{burshtein_photon-instanton_2021}. To find $f_k$, we evaluate the polynomial at the real frequencies of the array modes:
\begin{equation}
    f_k = f_k^{\textrm{apprx}} \times \sum_{l=0}^p \alpha_l (-\ii \omega_k)^l.
\end{equation}
Note that $f_k$ is real, since $\alpha_{1,3,\ldots}=0$. We have checked that the results are insensitive to the order $p$ if it is large enough (we used $p=6$), and that allowing for poles by fitting the brackets to a rational function (rather than a polynomial) does not significantly affect the extracted $f_k$ factors as well.

Before we proceed, let us stress that the expressions obtained in Subsections \ref{subsec:dS} and \ref{subsec:fk} are generic, and work for any nonlinear potential shape $V(\phi_0)$ with resonance frequency $\omega_0$, elastically-broadened by the environment with inverse $RC$ time $\Gamma_0$, and a UV-cutoff $v$. In the general case, we find
\begin{align}\label{eq:dS_generic}
    \delta S_1 =& \frac{1}{2}\sum_k \tilde{f}_k^2, \nonumber \\
    \tilde{f}_k =& \frac{1}{\pi}\sqrt{\frac{2\Delta\omega_k}{z}}\times \sqrt{\frac{\omega_k^2+\omega_0^2}{\Gamma_0\omega_k}\left|\delta\phi_0\left(\ii\omega_k\right)\right|^2+\left|\phi_0\left(\ii\omega_k\right)\right|^2}, \nonumber \\
    \delta S_2 =&\int_{-\infty}^{\infty}\id\tau\Big[V(\phi_0(\tau)) - V\left(\phi_0^{(0)}(\tau)\right) \nonumber \\
    &\hspace{0.7cm}- V^\p\left(\phi_0^{(0)}(\tau)\right)\delta\phi_0(\tau) -\frac{C_0 \omega_0^2}{2}\delta\phi_0^2(\tau)\Big],
\end{align}
where $V\left(\phi_0\right)$ is the noninverted potential and $C_0$ is the effective mass near the minima (where $V(\phi_0)\approx C_0 \omega_0^2\phi_0^2/2$), and the first line of Eq.~\eqref{eq:fk} still applies. All of these expressions may be obtained using the numerical solutions for $\phi_0^{(0)}$ (the trajectory of an instanton in the potential $V(\phi_0)$ decoupled from the environment) and $\phi_0$, $\delta\phi_0 = \phi_0 - \phi_0^{(0)}$. Furthermore, for a generic potential with resonance frequency $\omega_0$, we expect a first-order pole for $\phi_0^{(0)}(z)$ at $z = \omega_0$, which can be treated manually in Eq.~\eqref{eq:fk} by pulling a factor of $1/(\omega_0^2+\omega^2)$ out of the brackets and setting $\omega \rightarrow -\ii\omega_k$, leading to
\begin{align}
    f_k =&\frac{1}{\pi}\sqrt{\frac{2\Delta\omega_k}{z}}\frac{1}{\sqrt{\left(\omega_0^2 - \omega_k^2\right)^2 + \left(\Gamma_0\omega_k\right)^2}}\nonumber \\
    &\times\lim_{\omega\rightarrow -\ii\omega_k} \left[\left(\omega^2 +\Gamma_0 |\omega| +\omega_0^2\right) \phi_0(\ii\omega)\right],
\end{align}
where we used Eq.~\eqref{eq:delta_k}. Now, note that the approximate Eq.~\eqref{eq:phi0_apprx} holds regardless of the potential shape; again, anticipating the analytical structure of the full solution to be close to that of the approximated solution, we find that the expression inside the brackets should have no poles and can be analytically-continued using Eq.~\eqref{eq:phi0_cont}.

\subsection{Decay rate}
Once $f_k$ and $\delta S=\delta S_1 + \delta S_2$ have been obtained, all that is left to do is follow the same steps of Ref.~\cite{burshtein_photon-instanton_2021} to find the decay rate. Here we only cite the final result, and retrace the derivation in Appendix \ref{app:Tmat} for completeness. Summing over the squared elements of the $\calT$-matrix, given by Eq.~\eqref{eq:Tmat2} for all incoming and outgoing states, and accounting for the occupation of thermal photons at finite temperature, leads to
\begin{widetext}
\begin{align}\label{eq:Gink_sum}
    \Gamma_k^{\textrm{in}} =& \lambda_0^2 f_k^2 e^{-2\delta S} e^{-2\sum_{k^\p} f_{k^\p}^2 n_B(\omega_{k^\p})} \sum_{\Nout,\Nin} \sum_{\substack{k_1 < \ldots < k_{\Nout}\\k_1^\p < \ldots < k_{\Nin}^\p}} \prod_{j=1}^{\Nout} f_{k_j}^2\left(1+n_B(\omega_{k_j})\right)\prod_{j=1}^{\Nin} f_{k_j^\p}^2 n_B(\omega_{k_j^\p}) \nonumber \\
    &\times 2\pi \left[\delta\left(\omega_k + \sum_{j=1}^{\Nin}\omega_{k_j^\p} - \sum_{j=1}^{\Nout}\omega_{k_j}\right) - \{\omega_k \rightarrow -\omega_k\}\right],
    \end{align}
where $\Nout + \Nin\ge 3$ is odd, and the first and second delta functions correspond to absorption and emission of a photon at mode $k$, respectively. The sums may be rearranged as Taylor expansions of hyperbolic sines and cosines, yielding a compact expression for the decay rate, $\Gamma_k^{\textrm{in}} = 2f_k^2\Im \Pi_R(\omega_k)$, where the retarded self-energy of the photons is given by
\begin{equation}\label{eq:Gink}
        \Pi_R(\omega) = -\lambda_0^2 e^{-2\delta S}\int_0^{\infty}\id t\sin(\omega t) \exp\left(-\sum_{k^\p}\left[f_{k^\p}^2\left\{\left[1+n_B(\omega_{k^\p})\right]\left(1 - e^{-\ii\omega_{k^\p}t}\right) + n_B(\omega_{k^\p})\left(1 - e^{\ii\omega_{k^\p}t}\right)\right\} - f_{k^\p}^2\right]\right).
    \end{equation}
\end{widetext}
Furthermore, since Eq.~\eqref{eq:Tmat2} grants access to all elements of the $\calT$-matrix, it is possible to write refined rates, corresponding to higher-order response functions, such as the inelastic decay spectrum, which quantifies the production rate of a specific mode $k^\p$ in the decay process of a mode $k$. Again, the expressions are identical to those of Ref.~\cite{burshtein_photon-instanton_2021}, with the new $\delta S$ and $f_k$.

\section{Results} \label{sec:results}
\begin{figure}[b]
    \centering
    \includegraphics[width=0.98\columnwidth]{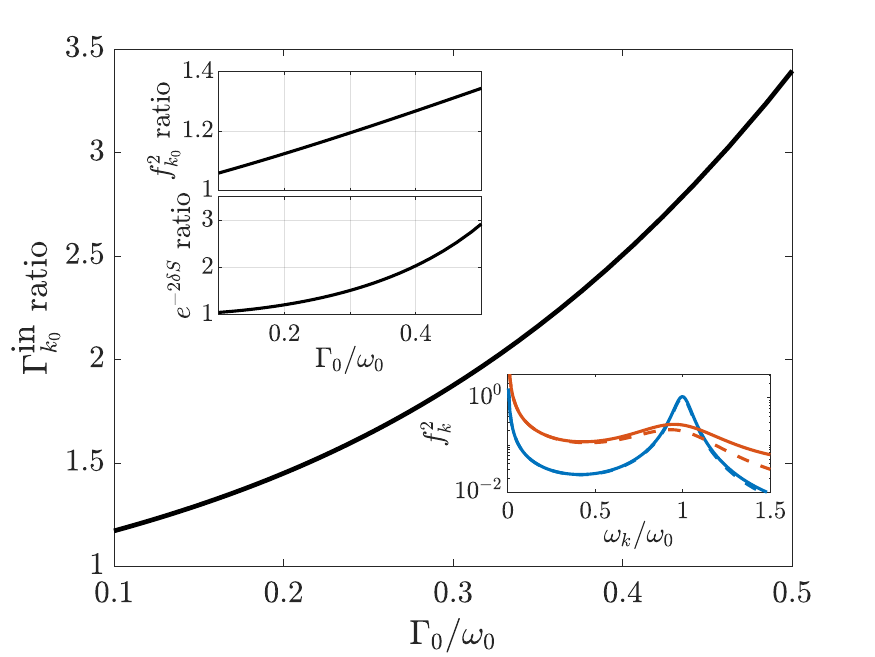}
    \caption{Main panel: Ratio between the on-resonance decay rate of this work to the on-resonance decay rate of Ref.~\cite{burshtein_photon-instanton_2021}, as function of $\Gamma_0/\omega_0$. Insets on top left: Ratios of action factors $e^{-2\delta S}$ and on-resonance $f_k^2$ factors. Inset on bottom right: $f_k^2$ factors of this work (continuous lines) and of Ref.~\cite{burshtein_photon-instanton_2021} (dashed lines) as function of $\omega_k/\omega_0$, for $\Gamma_0/\omega_0 = 0.1$ (blue) and $0.5$ (orange). In all plots, we the following experimentally-relevant parameters: $\omega_0 = 8$ GHz, $E_C=1.6$ GHz, $T=0$, and $z$ varies from $0.5$ to $3$.}
    \label{fig:onres_comp}
\end{figure}

The decay rate follows from a numerical evaluation of the imaginary part of Eq.~\eqref{eq:Gink}. Our results apply at energies above the emergent instanton scale, $\lambda^{\star} \ll \max\left\{\omega_k,T,\Gamma_0\right\}$, and assuming $E_C \ll E_J$ and $T \ll \omega_0$, such that the resonance frequency $\omega_0$ is not smeared by quantum or thermal fluctuations. We further require that the inelastic scattering rate $\Gamma_k^{\textrm{in}}/\Delta$ is much smaller than unity to justify the single-instanton approximation in Eq.~\eqref{eq:Smat}. Importantly, the numerical approach employed in this work relaxes the condition $\Gamma_0/\omega_0 \ll 1$, and extends the validity regime of the calculation to shorter $RC$ times compared to those of Ref.~\cite{burshtein_photon-instanton_2021}. Note, however, that at large enough $\Gamma_0/\omega_0$ and $E_C/E_J$, one has to consider contributions from scattering processes involving two or more instantons in Eq.~\eqref{eq:Smat}, as well as from non-Gaussian fluctuations in the path integral. In fact, the expansion around the minima of the transmon Josephson cosine term becomes less justified. All these contributions are beyond the scope of this work.

The $f_k^2$ factor in $\Gamma_k^{\textrm{in}} = 2f_k^2\Im \Pi_R(\omega_k)$ leads to a Lorentzian shape in $\omega_k$, centered around the resonance frequency $\omega_0$ with a width set by $\Gamma_0$. Here we focus on the on-resonance decay rate, $\omega_k = \omega_0$, which we denote by $\Gamma_{k_0}^{\textrm{in}}$. Fig.~\ref{fig:onres_comp} shows the on-resonance ratio between the decay rate of this work and the decay rate of Ref.~\cite{burshtein_photon-instanton_2021} as a function of $\Gamma_0/\omega_0$, for experimentally-relevant parameters (listed in the caption). As expected, the ratio approaches unity as $\Gamma_0/\omega_0 \rightarrow 0$, where the approximations of Ref.~\cite{burshtein_photon-instanton_2021} apply. The numerical solution enhances the decay rate at larger $\Gamma_0/\omega_0$. As illustrated by the insets of Fig.~\ref{fig:onres_comp}, this enhancement is due to both a larger on-resonance $f_k$ factor, as well as a smaller suppression of the action, $e^{-2\delta S}$.

The on-resonance decay rates for several transmon-array devices, whose parameters are listed in Table \ref{tab:devs}, were measured in Ref.~\cite{kuzmin_inelastic_2021}. A simplified version of the instanton calculation of Ref.~\cite{burshtein_photon-instanton_2021} provided predictions that agreed nicely with the measurements of devices with $z > 1$, but fell short of the measured decay rates of $z < 1$ devices, where the $RC$ relaxation times are not much larger than the tunneling time. The results of this work reasonably agree with all devices measured in Ref.~\cite{kuzmin_inelastic_2021} (other than 4a, see below), as shown in Fig.~\ref{fig:exp_comp}. The theory lines are broadened by a $\pm10\%$ uncertainty in the values of $E_C$ and $Z$, typical in such experiments (see the supplemental material in Ref.~\cite{kuzmin_inelastic_2021} for a discussion of the transmission line characterization and the associated uncertainties). A temperature of $T = 40$ mK, extracted from qubit experiments on similar setups, was assumed for all devices; in practice, temperature could vary between experiments, which could have a small but non-negligible effect on the results. To avoid overfitting the data using too large experimental deviations, we use the same temperature in all devices. To demonstrate the potential effect of different temperatures, we also show results for $T = 50$ mK in Appendix \ref{app:temp}. The Josephson energies $E_J$ were extracted from the measured $\omega_0$ and $E_C$, by inverting Eq.~\eqref{eq:w0_Mathieu_avg}. We stress that there are no fitting parameters.

We note that the new theory fails to capture the measurements of device 4a. Furthermore, the slopes of the theory lines (in logarithmic scale) deviate from the slopes of the experimental measurements for the $z<1$ devices at small resonance frequencies, where the ratios $\Gamma_0/\omega_0$ and $E_C/E_J$ are relatively large. As mentioned above, these deviations could be attributed to high-order effects beyond the scope of this work.

\begin{table}[t]
    \centering
    \begin{tabular}{|*{9}{P{15mm}|}}
         \hline
         Device&   $Z$ [k$\Omega$]& $z$&$E_C$ [GHz]&$\Gamma_0$ [GHz] \\\hline\hline
         1a& 
     5.26& 0.81&0.66& 1.03\\
 2a& 5.36& 0.83&1.11& 1.70\\
 3a& 4.44& 0.69&1.35& 2.50\\
 4a& 5.22& 0.81&1.96& 3.10\\
 1b& 9.25& 1.43&0.74& 0.66\\
 2b& 13.61& 2.11&1.05& 0.63\\
 3b& 10.89& 1.68&1.56& 1.18\\
 4b& 11.38& 1.76&1.83& 1.32\\ \hline \end{tabular}
    \caption{Nominal device parameters.}
    \label{tab:devs}
\end{table}

\begin{figure*}[t]
    \centering
    \includegraphics[width=0.98\textwidth]{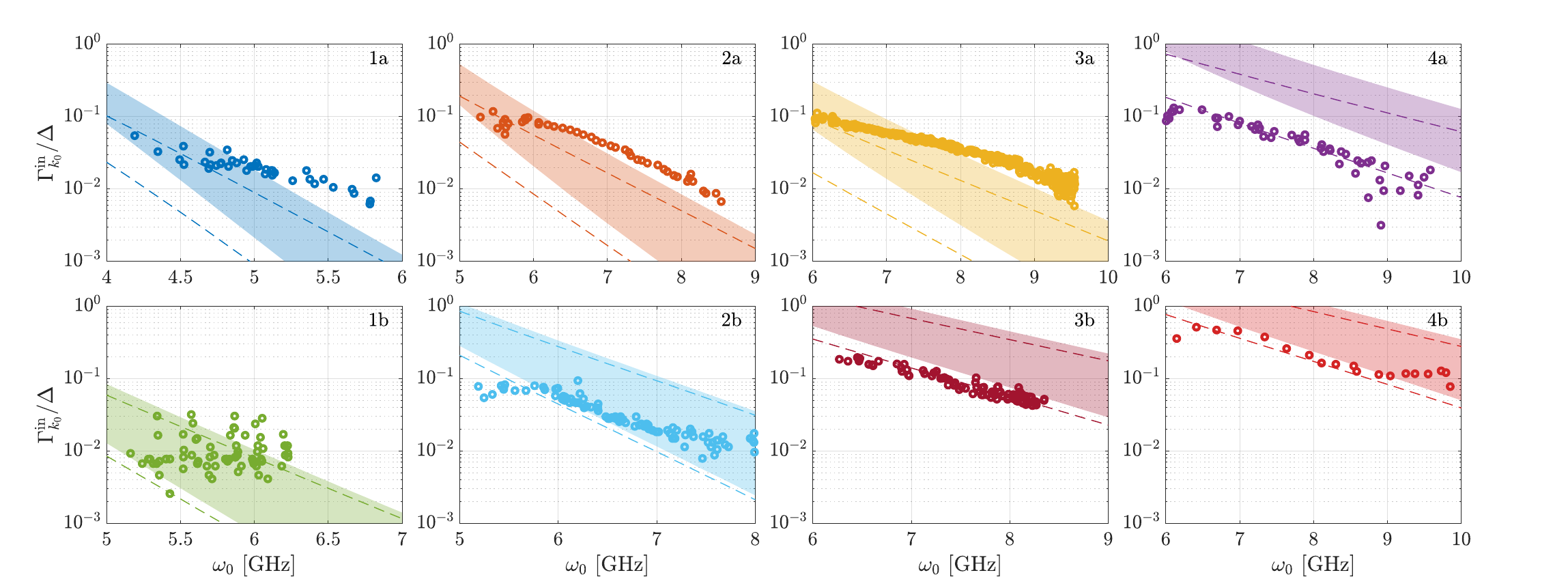}
    \caption{Comparison of the calculated on-resonance decay rate (shaded regions) with the experimental data from Ref.~\cite{kuzmin_inelastic_2021}, with no fitting parameters. The captions correspond to the devices listed in Table \ref{tab:devs}. We assume an uncertainty of $\pm 10\%$ in the values of $Z$ and $E_C$, broadening the theory lines. The dashed lines mark the upper and lower limits of the theory predictions of Ref.~\cite{burshtein_photon-instanton_2021}. A typical temperature of 40 mK was used in producing the theory lines.}
    \label{fig:exp_comp}
\end{figure*}

\section{Conclusions} \label{sec:conclusions}
We developed a numerical scheme for 
finding the real-time photon-instanton cross-section in a quadratic transmission line coupled to a degenerate, nonlinear potential. Our work relies on the analytical formalism developed in Ref.~\cite{burshtein_photon-instanton_2021}, that showed that a photon in the transmission line can trigger an instanton which leads to the emission of multiple photons at lower-frequencies with order-unity probability. Here we extended the validity regime of the formalism to scenarios where the relaxation $RC$ time, proportional to the impedance of the transmission line, is not much larger than the tunneling time of the instanton, by solving the joint semi-classical instanton-photons trajectories numerically, thus lifting some of the approximations needed to obtain the analytical solution of Ref.~\cite{burshtein_photon-instanton_2021}. Our new predictions improve the agreement with the experiment in Ref.~\cite{kuzmin_inelastic_2021} without fitting parameters for devices that were not captured by the results of Ref.~\cite{burshtein_photon-instanton_2021}, particularly those whose impedance is smaller than the resistance quantum ($z = Z/R_Q < 1$). 



The joint semi-classical trajectories in imaginary time of the instanton at the impurity and the response of the array were determined by a set of $N + 1$ coupled differential equations. We exploited the linearity of the array, which is diagonalized by the bulk Fourier modes, to reduce the set of coupled equations to a single ordinary differential equation that may be solved iteratively, or, equivalently, to a single integro-differential equation. The $\calT$-matrix elements, determining the photon-instanton cross-section, were obtained by numerical analytic continuation of the solutions to the semi-classical equations of motion. We used our knowledge of the analytical structure of the approximate solution from Ref.~\cite{burshtein_photon-instanton_2021} to avoid the major issues associated with this ill-posed problem.


The formalism developed in this work is general, and could be applied to any degenerate nonlinear potential coupled to a linear environment. In particular, an analytical expression for the instanton solution isolated from the environment (denoted by $\phi_0^{(0)}$ in this work) is not necessary. Therefore, our method could be useful in many other field-theoretical contexts of condensed matter and high energy systems. This method could be also relevant to recent works on tunneling in fluxonium qubits and resonant Josephson junctions \cite{vakhtel_quantum_2023,vakhtel_tunneling_2024,randeria_dephasing_2024, mencia_integer_2024}.

\nocite{burshtein_zenodo16596445_2025}

\subsection*{Acknowledgements}
Our work has been supported by the ISF and the Directorate for Defense Research and Development (DDR\&D) Grant No. 3427/21, the ISF Grant No. 1113/23, and the US-Israel Binational Science Foundation (BSF) Grant No. 2020072. A.B. is also supported by the Adams Fellowship Program of the Israel Academy of Sciences and Humanities. 

\appendix

\section{Transmon charge dispersion}\label{app:trans}
\subsection{An isolated transmon}
Consider a transmon disconnected from the environment, with the Hamiltonian
\begin{equation}
    \mathcal{H}_0=E_C(Q_0 - q_g)^2-E_J\cos(2\phi_0),
\end{equation}
where $Q_0$ and $\phi_0$ are the charge and flux of the transmon, respectively, obeying $[\phi_0,Q_0]=\ii$, and $q_g$ is an offset charge imposed by e.g. stray capacitances, that can be gauged away in the limit of a semi-infinite transmission line, and is therefore ignored in the rest of this work. The Schr{\"o}dinger equation of this Hamiltonian can be recast to the form of the Mathieu equation \cite{abramowitz_handbook_1965, koch_charge-insensitive_2007}, yielding the exact eigenenergies $E_m(q_g)$ in terms of the Mathieu characteristic values, $a_\nu(E_J/(2E_C))$ and $b_\nu(E_J/(2E_C))$. The bands are periodic in $q_g$, $E_m(q_g \pm 2) = E_m(q_g)$. The resonance frequency of the transmon is given by $\omega_0(q_g)=E_1(q_g)-E_0(q_g)$, and generally depends on $q_g$. In the limit of small $E_C/E_J$, we recover $\omega_0 \approx \sqrt{8 E_J E_C}$, which corresponds to the expansion of $\cos(2\phi_0)\approx 1 - 2\phi_0^2$ to harmonic order. Corrections to $\omega_0$ can be derived systematically in perturbation theory; the leading correction in $E_C/E_J$, stemming from the quartic term in the expansion of $\cos(2\phi_0)$, gives $\omega_0 \approx \sqrt{8 E_J E_C} - E_C$.

In the experiment \cite{kuzmin_inelastic_2021}, the resonance frequency $\omega_0$ and charging energy $E_C$ are measured directly, and the expression for $\omega_0$ in terms of the Mathieu characteristic values can be used to extract the Josephson energy $E_J$. Since $q_g$ can fluctuate, we take the average over its extreme values,
\begin{align}\label{eq:w0_Mathieu_avg}
    \omega_0 =& \frac{1}{2}\left(\omega_0(q_g = 1) + \omega_0(q_g = 0)\right) \nonumber \\
    =&\frac{E_C}{2}\left(b_2(\chi) - a_0(\chi) + a_1(\chi) - b_1(\chi)\right),
\end{align}
where $\chi = E_J/(2E_C)$.

The dependence of $E_m$ on the quasicharge $q_g$ determines the width of the Bloch bands of the transmon. In this work we are only interested in the charge dispersion of the $m=0$ band,
\begin{align}
    \lambda_0 =& \frac{1}{2}\left(E_0(q_g=1) - E_0(q_g=0)\right) \nonumber \\
    =&\frac{E_C}{2}\left(b_1(E_J/2E_C) - a_0(E_J/2E_C)\right) \nonumber \\
    \approx& \frac{8}{\sqrt{\pi}}(8E_J^3E_C)^{1/4}e^{-\sqrt{8E_J/E_C}},
\end{align}
where the last form follows from the WKB approximation.

\subsection{Renormalization of the charge dispersion}
The charge dispersion $\lambda_0$ is renormalized upon connecting the transmon to the transmission line. This is evident from Eq.~\eqref{eq:Tmat2}, where the back-action of the array modes in response to the instanton leads to a renormalization $\lambda_0 \rightarrow e^{-\delta S} \lambda_0$. To find the low-energy fixed point of the transmon, one may derive renormalization group (RG) flow equations; note that at large $E_J/E_C$, the lowest energy band of the periodic transmon potential is approximately given by $\lambda_0 \cos(\pi q_g)$, such that the flow equations are dual to those of the boundary sine-Gordon model \cite{gogolin_bosonization_2004} (i.e.~$\cos(2\phi_0)\rightarrow\cos(\pi q_g)$, and $z\rightarrow 1/z$). One then finds that instantons are relevant in an RG sense for $z>1$, and the transmon becomes a phase slip junction at low energies, as predicted by the Schmid-Bulgadaev quantum phase transition \cite{schmid_diffusion_1983, bulgadaev_phase_1984}. The RG flow at $z > 1$ gives rise to a characteristic energy scale $\lambda^{\star}\sim(\lambda_0/\Lambda^{1/z})^{1/(1 - 1/z)}$ ($\Lambda$ is a high-energy cutoff that can be interpreted as the transmon resonance frequency, $\Lambda\sim\omega_0$), at which the renormalized $\lambda_0$ becomes comparable with the high-energy cutoff. At energies below $\lambda^{\star}$, the transmon becomes a phase slips junction, the resonance frequency $\omega_0$ is irrelevant, and the approximations and assumptions discussed above are no longer valid --- for instance, there is no justification to expand $\cos(2\phi_0)$ around $\phi_0 = 0$ in Eq.~\eqref{eq:LagSys}. In other words, instanton effects are nonperturbative. In this work we consider only the perturbative instanton regime, where $\max\left\{\omega_k,T,\Gamma_0\right\} \gg \lambda^{\star}$.

\section{Integro-differential equation for $\phi_0$}
\label{app:kernel}
To get Eq.~\eqref{eq:integro-diff-kernel}, we start with the Euler-Lagrange equation for $\phi_0$ in Eq.~\eqref{eq:full_eq_system} written in terms of the bulk modes:
\begin{equation}
    \label{eq:EL-phi0-bulk}
    \partial_{\tau}^2\phi_0=\frac{\omega_0^2}{2}\sin(2\phi_0)+v\Gamma_0\phi_0-v\Gamma_0\sum_q{\sin{\hat{\delta}_q}\hat{\phi}_q}.
\end{equation}
Plugging the result of Eq.~\eqref{eq:bulk-phi0}, transformed to the $\tau$-domain, into the above, we get an integro-differential equation for $\phi_0$:
\begin{align}
    \label{eq:integro-diff-explicit}
    \partial_{\tau}^2\phi_0(\tau)=&\frac{\omega_0^2}{2}\sin(2\phi_0(\tau))+v\Gamma_0\phi_0(\tau)\nonumber\\
    &\hspace{-1cm}-\frac{v^2\Gamma_0}{Z}\sum_q\frac{\sin^2{\hat{\delta}_q}}{C_q}\int_{-\infty}^\infty{\frac{\id\omega}{2\pi}e^{-\ii\omega\tau}\frac{\phi_0(\omega)}{\omega^2+\omega_q^2}}.
\end{align}
The integral can be represented as a convolution in imaginary time. By introducing the kernel
\begin{equation}
    \label{eq:kernel}
    K(\tau)=\frac{v^2\Gamma_0}{2Z}\sum_q\frac{\sin^2{\hat{\delta}_q}}{\omega_q C_q}e^{-\omega_q|\tau|},
\end{equation}
Eq.~\eqref{eq:integro-diff-explicit} simplifies to \eqref{eq:integro-diff-kernel}. To get an explicit expression for the kernel, we evaluate the sum over $q$ as an integral in the thermodynamic limit, $\sum_q \rightarrow (N/\pi) \int_0^\pi \id q$:
\begin{align}
\label{eq:kernel_full_derivation}
    K(\tau)=&\frac{v^2\Gamma_0}{2Z}\int_0^{\pi}{\frac{\id q}{\pi/N}\frac{\sin^2{\hat{\delta}_q}}{\omega_q C_q}e^{-\omega_q|\tau|}}\nonumber \\
    =&\frac{v^3\Gamma_0}{\pi}\int_0^{2v}\frac{\id \omega_q}{v\omega_q}\frac{\omega_q^2}{v^2}\sqrt{1-\left(\frac{\omega_q}{2v}\right)^2}e^{-\omega_q|\tau|} \nonumber \\
    =&v\Gamma_0\left(\frac{L_2(2v|\tau|)-I_2(2v|\tau|)}{|\tau|}+\frac{4v}{3\pi}\right),
\end{align}
where $L_2(x)$ is the modified Struve  function and $I_2(x)$ is the modified Bessel function of the first kind \cite{abramowitz_handbook_1965}. For $v|\tau|\gg1$ we have
\begin{equation}
\label{eq:kernel_approx}
    K(|\tau|\gg1/v)\approx\frac{4v^2\Gamma_0}{\pi}\int_0^1{\id x xe^{-2v|\tau|x}}=\frac{\Gamma_0}{\pi\tau^2}.
\end{equation}

Note that we are looking for a solution $\phi_0$ that jumps by $\pi$. It is convenient to write $\phi_0 = \phi_0^{(0)}+\delta\phi_0$ and reformulate Eq.~\eqref{eq:integro-diff-kernel} as an equation for $\delta\phi_0$:
\begin{align} \label{eq:integro-diff-dphi0}
    \partial_{\tau}^2 \delta\phi_0 =& \frac{\omega_0^2}{2}\left[\sin\left(2\phi_0^{(0)}+2\delta\phi_0\right) - \sin\left(2\phi_0^{(0)}\right)\right] \nonumber \\
    &\hspace{-1cm}+ v\Gamma_0\delta\phi_0 - \delta\phi_0(\tau) * K(\tau) + F\left(\phi_0^{(0)}(\tau)\right),
\end{align}
where $F\left(\phi_0^{(0)}(\tau)\right)$ is the inverse Fourier transform of $F\left(\phi_0^{(0)}(\ii\omega)\right) = \Gamma_0 |\omega|\phi_0^{(0)}(\ii\omega)$. Although mathematically equivalent, the numerical solution of Eq.~\eqref{eq:integro-diff-dphi0} is easier than that of Eq.~\eqref{eq:integro-diff-kernel}, since the solution $\delta\phi_0$ satisfies $\delta\phi_0(\tau\rightarrow\pm\infty)=0$.

\section{Linearized equations of motion}\label{app:PRL_lim}
In this Appendix, we recap the approximations applied in Ref.~\cite{burshtein_photon-instanton_2021}, and show that the results of this work agree with those of Ref.~\cite{burshtein_photon-instanton_2021} in the appropriate limits.

Consider the Lagrangian expressed in terms of the bulk modes of the array without the transmon, given by the Wick-rotated version of Eq.~\eqref{eq:LagBulk}. Let us write $\phi_0 = \phi_0^{(0)} + \delta\phi_0$, and expand the Lagrangian around the solution to the isolated transmon, $\phi_0^{(0)}$, to second order in $\delta\phi_0$:
\begin{align}\label{eq:L_linear}
    \mathcal{L} =& \frac{C_0}{2}\left(\partial_{\tau}\phi^{(0)}_0\right)^2 - E_J\cos\left(2\phi^{(0)}_0\right)\nonumber \\
    &+\frac{C_0}{2}\left(\partial_{\tau}\delta\phi_0\right)^2 + 2E_J\cos\left(2\phi_0^{(0)}\right)\delta\phi_0^2 \nonumber \\
    &+ \frac{1}{2L}\left[2\phi_0^{(0)}\left(\delta\phi_0-\phi_1\right) + \delta\phi_0^2 - 2\delta\phi_0\phi_1\right] \nonumber \\
    &+\sum_q\left[\frac{C_q}{2}\left(\partial_{\tau}\hat{\phi}_q\right)^2 + \frac{C_q\omega_q^2}{2}\hat{\phi}_q^2\right],
\end{align}
where we used the equation that $\phi_0^{(0)}$ satisfies, $\partial_{\tau}^2\phi_0^{(0)} = \left(\omega_0^2/2\right)\sin\left(2\phi_0^{(0)}\right)$, to discard a complete derivative term. The Euler-Lagrange equation for $\delta\phi_0$ reads
\begin{equation}
    \partial_{\tau}^2\delta\phi_0 = \omega_0^2 \cos\left(2\phi_0^{(0)}\right)\delta\phi_0 + v\Gamma_0\left(\phi_0^{(0)}+\delta\phi_0-\phi_1\right).
\end{equation}
Following the approximations of Ref.~\cite{burshtein_photon-instanton_2021}, we replace $\cos\left(2\phi_0^{(0)}\right)$ with unity. That is, we replace $\cos\left(2\phi_0^{(0)}\right)$ with its value at the minima of the potential, ignoring the trajectory that $\phi_0^{(0)}$ traverses between two minima over a time $\sim 1/\omega_0$. We then find
\begin{equation}\label{eq:dphi0_apprx_eq}
    \partial_{\tau}^2\delta\phi_0 = \omega_0^2 \delta\phi_0 + v\Gamma_0\left(\phi_0^{(0)}+\delta\phi_0-\phi_1\right).
\end{equation}
Let us stress that Eq.~\eqref{eq:dphi0_apprx_eq} holds for any potential shape $V(\phi_0)$, not necessarily sinusoidal, upon expanding the Lagrangian to second order in $\delta\phi_0$ and replacing $V^{\p\p}\left(\phi_0^{(0)}\right)$ with its value at minimum, $V^{\p\p}\left(\phi_0^{(0)}\right)\rightarrow C_0 \omega_0^2$.
The solution for $\delta\phi_0$ in Matsubara frequency follows easily, using Eq.~\eqref{eq:phi_1-phi_0}; we find
\begin{equation}\label{eq:dphi0_apprx}
    \delta\phi_0^{\textrm{apprx}}(\ii\omega) = -\frac{\Gamma_0|\omega|}{\omega^2 + \Gamma_0|\omega| + \omega_0^2}\phi_0^{(0)}(\ii\omega).
\end{equation}
This solution leads to Eq.~\eqref{eq:phi0_apprx}, which was used in Section \ref{sec:Tmat}.

One may also find an analytical solution for the $k$ modes of the full system in terms of $\phi_0^{(0)}$. Let us write $\phi_n = \delta\phi_n^{(0)} + \delta\phi_n$, with $\phi_0^{(0)}$ given by the trajectory of an isolated instanton and $\phi_n^{(0)}=0$, and define $\delta\phi_n = \sum_k\delta\phi_k\sin(kn+\delta_k)$. Within the same approximations, the Lagrangian \eqref{eq:L_linear} may be written as
\begin{align}\label{eq:L_linear_k}
    \mathcal{L} =& \frac{C_0}{2}\left(\partial_{\tau}\phi^{(0)}_0\right)^2 - E_J\cos\left(2\phi_0^{(0)}\right) +\frac{1}{2L}\left(\phi_0^{(0)}\right)^2 \nonumber \\
    &+\sum_k \left[\frac{C_k}{2} \left(\partial_{\tau}\delta\phi_k\right)^2 +\frac{C_k\omega_k^2}{2}\delta\phi_k^2\right] \nonumber \\
    &-\frac{\sin\left(k+\delta_k\right)-\sin\delta_k}{L}\phi_0^{(0)}\delta\phi_k.
\end{align}
Solving the Euler-Lagrange equations for $\delta\phi_k$, we get Eq.~\eqref{eq:phik_apprx},
\begin{equation}
    \delta\phi_k^{\textrm{apprx}}(\ii\omega) = \frac{\omega_k\cos\delta_k}{ZC_k}\frac{1}{\omega^2+\omega_k^2}\phi_0^{(0)}(\ii\omega).
\end{equation}

The action obtained by plugging $\delta\phi_k^{\textrm{apprx}}$ to the Lagrangian in Eq.~\eqref{eq:L_linear_k} was calculated in Ref.~\cite{burshtein_photon-instanton_2021}, and is given by $S = S_0 + \delta S^{\textrm{apprx}}$, with $S_0 = \sqrt{8 E_J/E_C}$ and $\delta S^{\textrm{apprx}} = \sum_k \left(\tilde{f}_k^{\textrm{apprx}}\right)^2/2$, where $\tilde{f}_k^{\textrm{apprx}} = \sqrt{2\Delta/(z\omega_k)}/\cosh(\pi\omega_k/(2\omega_0))$. As a sanity check, we show that the action $\delta S = \delta S_1 + \delta S_2$ of Eq.~\eqref{eq:dS} converges to $\delta S^{\textrm{apprx}}$ in the appropriate limits. First, consider the $\tilde{f}_k$ factor, given by Eq.~\eqref{eq:fkTilde}; using Eq.~\eqref{eq:dphi0_apprx}, we find
\begin{equation}
    \tilde{f}_k \approx \tilde{f}_k^{\textrm{apprx}} \times \sqrt{\frac{\omega_k^2 + \omega_0^2}{\omega_k^2 + \Gamma_0 \omega_k + \omega_0^2}}.
\end{equation}
To leading order in $\Gamma_0/\omega_k$, the square root in the expression above can be replaced with unity, leading to $\tilde{f}_k \rightarrow \tilde{f}_k^{\textrm{apprx}}$. For $\delta S_2$, expanding $\cos(2\phi_0)$ around $\phi_0^{(0)}$ to second order in $\delta\phi_0$ and replacing $\cos\left(2\phi_0^{(0)}\right)$ with unity, we find
\begin{equation}
    \delta S_2 \approx -\frac{v}{2\Gamma_0}\int_{-\infty}^{\infty}\id\tau\delta\phi_0^2(\tau)\left[1 - \cos\left(2\phi_0^{(0)}(\tau)\right)\right]\rightarrow 0.
\end{equation}
Overall, we get $\delta S \rightarrow \sum_k \Delta/(z\omega_k)\times 1/\cosh^2(\pi\omega_k/(2\omega_0))$, in agreement with Ref.~\cite{burshtein_photon-instanton_2021}.

\section{Corrections to $\phi_0(\ii\omega)$ using the asymptotic behavior of $\phi_0(|\tau|\gg1/\omega_0)$}\label{app:phi0}
Let us refer to a technical point in the evaluation of $\phi_0(\ii\omega)$ that is necessary to ensure stable analytical continuation using Eq.~\eqref{eq:phi0_cont}. An issue arises due to the slow decay of the solution at large $|\tau|$, $\delta\phi_0(|\tau| \gg 1/\omega_0)\sim 1/\tau$, shown in Fig.~\ref{fig:phi0_time}. This asymptotic behavior can be inferred from the approximate solution --- from the inverse Fourier transform of Eq.~\eqref{eq:dphi0_apprx}, we find $ \delta\phi_0^{\textrm{apprx}}(|\tau| \gg 1/\omega_0)\sim 1/\tau$. At those large times, the numerical solution $\delta\phi_0(\tau)$ should approach $\delta\phi_0^{\textrm{apprx}}(\tau)$; this is evident in Fig.~\ref{fig:phi0_time}, and follows from the approximation
\begin{equation}
    \sin\left(2\phi_0^{(0)}(\tau) + 2\delta\phi_0(\tau)\right) - \sin\left(2\phi_0^{(0)}(\tau)\right) \approx 2\delta\phi_0(\tau),
\end{equation}
used to derive the approximated Eq.~\eqref{eq:dphi0_apprx_eq}, which holds for $|\tau| \gg 1/\omega_0$. That is, at large $|\tau|$ the equation for $\delta\phi_0(\tau)$ reduces to that of $\delta\phi_0^{\textrm{apprx}}(\tau)$, and thus $\delta\phi_0(\tau)\sim\delta\phi_0^{\textrm{apprx}}(\tau)\sim 1/\tau$.

Now, the numerical solution for $\delta\phi_0(\tau)$ is obtained over a finite support of imaginary times, $|\tau| \le \tau_{\max}$, which effectively truncates $\delta\phi_0(\tau)$ and discards a potentially important contribution due to the slow asymptotic decay. This can be easily fixed by extrapolating $\delta\phi_0(\tau)$ to $|\tau| > \tau_{\max}$ using the $1/\tau$ behavior (assuming $\tau_{\max} \gg 1/\omega_0$); namely,\begin{equation}
    \delta \phi_0(\tau > \tau_{\max}) = \delta \phi_0(\tau_{\max}) \times \frac{\tau_{\max}}{\tau},
\end{equation}
and similarly for $\tau < -\tau_{\max}$ (recall that $\delta\phi_0(-\tau) = -\delta\phi_0(\tau)$). Then,
\begin{align}
    \delta\phi_0(\ii\omega) \approx& \int_{-\tau_{\max}}^{\tau_{\max}}\id\tau e^{\ii\omega\tau} \delta\phi_0(\tau) \nonumber \\
    &\hspace{0.5cm}+ 2\int_{\tau_{\max}}^\infty\id\tau e^{\ii\omega\tau} \delta \phi_0(\tau_{\max}) \times \frac{\tau_{\max}}{\tau} \nonumber \\
    =& \int_{-\tau_{\max}}^{\tau_{\max}}\id\tau e^{\ii\omega\tau} \delta\phi_0(\tau) \nonumber \\
    &\hspace{0.5cm}+ 2\tau_{\max}\delta\phi_0(\tau_{\max})\Gamma(0,-\ii\tau_{\max}\omega),
\end{align}
where $\Gamma(s,x)$ is the incomplete gamma function \cite{abramowitz_handbook_1965}. The first term above is computed numerically. Failing to correct it using the second term could lead to non-negligible oscillations in Matsubara frequency that would be detrimental to the fit in Eq.~\eqref{eq:phi0_cont}.

\section{Details of the derivation and evaluation of the inelastic decay rate}\label{app:Tmat}
For completeness, we retrace the steps of Ref.~\cite{burshtein_photon-instanton_2021} to derive the inelastic decay rates of the photons from the elements of the $\calT$
-matrix.

\subsection{Derivation of Eq.~\eqref{eq:Gink}}
The elements of the $\calT$-matrix yield the rates of the inelastic scattering processes, by means of the Fermi golden rule. In the following, we assume that the thermodynamic limit holds, such that the set of outgoing states forms a continuous bath for the decay of a photon of mode $k$.

First, before taking the squared absolute value, one must sum coherently the contributions of instantons and anti-instantons in Eq.~\eqref{eq:Tmat2}. This eliminates the contribution of terms with odd $\Nout + \Nin$, and multiplies by 2 the contribution of terms with even $\Nout+\Nin$, due to the parity symmetry $\phi_n\rightarrow -\phi_n$ (for the entire array) of the Lagrangian in Eq.~\eqref{eq:LagSys}.

Consider a process with $\Nin + 1$ incoming photons and $\Nout$ outgoing photons (with odd $\Nout + \Nin\ge 3$), in which a photon at a specific mode $k$ collides with photons at modes $k_1^\p,\ldots,k_{\Nin}^\p$ to produce photons at $k_1,\ldots,k_{\Nout}$. The rate of such a process is given by the Fermi golden rule,
\begin{align}\label{eq:Gink_process}
    \Gamma_{k_1,\ldots,k_{\Nout}}^{k,k_1^\p,\ldots,k_{\Nin}^\p} =& \lambda_0^2 f_k^2 e^{-2\delta S} \prod_{j=1}^{\Nout}f_{k_j}^2\prod_{j=1}^{\Nin}f_{k_j^\p}^2 \nonumber \\
    &\times 2\pi\delta\left(\omega_k + \sum_{j=1}^{\Nin}\omega_{k_j^\p} - \sum_{j=1}^{\Nout}\omega_{k_j}\right).
\end{align}
Note that photons may be added to both the incoming and outgoing states without changing the physical scattering process, but this would lead to a different $\calT$-matrix element. For example, a single photon at mode $k^{\p\p}$ may be added to both incoming and outgoing photons to produce a physically-equivalent scattering process, but its corresponding $\calT$-matrix element would be multiplied by $-f_{k^{\p\p}}^2$. Summing the amplitude of the scattering processes with and without the $k^{\p\p}$ photons and then squaring the result, we find that when a photon at $k^{\p\p}$ is present in the system, the corresponding rate would be multiplied by a factor $\left(1 - f_{k^{\p\p}}^2\right)^2$. At finite temperatures, the occupation of this mode is determined by the Bose-Einstein distribution, $n_B(\omega_{k^{\p\p}})$, such that the overall factor multiplying the rate is
\begin{equation}
    \left[1-n_B(\omega_{k^{\p\p}})\right] + \left(1 - f_{k^{\p\p}}^2\right)^2n_B(\omega_{k^{\p\p}}) \approx e^{-2f_{k^{\p\p}}^2n_B(\omega_{k^{\p\p}})},
\end{equation}
where the approximation holds since $f_{k^{\p\p}}^2\sim 1/N$ (as can be seen from Eqs.~\eqref{eq:fk_apprx} and \eqref{eq:fk}). Similarly, since the $f_k$ factors are very small in the thermodynamic limit, it is unnecessary to explicitly consider cases in which some mode appears more than once in the incoming or outgoing states.

The total decay rate of an incoming photon at mode $k$ is found by summing over the rates of the form of Eq.~\eqref{eq:Gink_process}, multiplied by a factor $e^{-2\sum_{k^{\p\p}} f_{k^{\p\p}}^2 n_B(\omega_{k^{\p\p}})}$ following the discussion above, and subtracting the rates of processes in which the photon at $k$ is emitted rather than absorbed. Furthermore, one has to multiply the rates by factors of $1+n_B(\omega_{k_j})$ for each of the outgoing photons in Eq.~\eqref{eq:Gink_process} to account for spontaneous and stimulated emission, and by factors of $n_B(\omega_{k_j^\p})$ for each of the incoming photons to account for spontaneous absorption. One arrives at Eq.~\eqref{eq:Gink_sum},
\begin{widetext}
    \begin{align}
    \Gamma_k^{\textrm{in}} =& \lambda_0^2 f_k^2 e^{-2\delta S} e^{-\sum_{k^\p} 2 f_{k^\p}^2 n_B(\omega_{k^\p})} \sum_{\Nout,\Nin} \sum_{\substack{k_1 < \ldots < k_{\Nout}\\k_1^\p < \ldots < k_{\Nin}^\p}} \prod_{j=1}^{\Nout} f_{k_j}^2\left(1+n_B(\omega_{k_j})\right)\prod_{j=1}^{\Nin} f_{k_j^\p}^2 n_B(\omega_{k_j^\p}) \nonumber \\
    &\times 2\pi \left[\delta\left(\omega_k + \sum_{j=1}^{\Nin}\omega_{k_j^\p} - \sum_{j=1}^{\Nout}\omega_{k_j}\right) - \{\omega_k \rightarrow -\omega_k\}\right],
    \end{align}
where $\Nout + \Nin\ge 3$ is odd.

    We proceed by writing the energy-conserving delta functions using their Fourier representation, $2\pi\delta(\omega) = \Re\left\{\int_0^{\infty} \id t e^{\ii\omega t}\right\}$. This allows us to rewrite $\Gamma_k^{\textrm{in}}$ as
    \begin{equation} \label{eq:Gink_inter}
        \Gamma_k^{\textrm{in}} = 2\lambda_0^2 f_k^2 e^{-2\delta S} e^{-\sum_{k^\p} 2 f_{k^\p}^2 n_B(\omega_{k^\p})} \Re \left\{\int_{0}^{\infty}\id t \left(e^{\ii\omega_k t} - e^{-\ii\omega_k t}\right)\sum_{\Nout,\Nin}\frac{\sigma_{\textrm{out}}^{\Nout}(t)\sigma_{\textrm{in}}^{\Nin}(t)}{\Nout !\Nin !}\right\},
    \end{equation}
    where
    \begin{equation}
        \sigma_{\textrm{out}}(t) = \sum_{k^\p} f_{k^\p}^2 \left(1+n_B(\omega_{k^\p})\right)e^{-\ii\omega_{k^\p} t},\quad \sigma_{\textrm{in}}(t) = \sum_{k^\p} f_{k^\p}^2 n_B(\omega_{k^\p})e^{\ii\omega_{k^\p} t}.
    \end{equation}
    The sum over $\Nin, \Nout$ explicitly reads
    \begin{align}\label{eq:sigma_sum}
    \sum_{\Nout,\Nin}\frac{\sigma_{\textrm{out}}^{\Nout}(t)\sigma_{\textrm{in}}^{\Nin}(t)}{\Nout !\Nin !} =& \sum_{\Nout = 1,3,\ldots}\frac{\sigma_{\textrm{out}}^{\Nout}(t)}{\Nout !} \sum_{\Nin = 2,4,\ldots}\frac{\sigma_{\textrm{in}}^{\Nin}(t)}{\Nin !} + \sum_{\Nout = 3,5,\ldots}\frac{\sigma_{\textrm{out}}^{\Nout}(t)}{\Nout !} \nonumber \\
    &+\sum_{\Nout = 2,4,\ldots}\frac{\sigma_{\textrm{out}}^{\Nout}(t)}{\Nout !} \sum_{\Nin = 1,3,\ldots}\frac{\sigma_{\textrm{in}}^{\Nin}(t)}{\Nin !} + \sum_{\Nin = 3,5,\ldots}\frac{\sigma_{\textrm{in}}^{\Nin}(t)}{\Nin !} \nonumber \\
    =& \sinh\left[\sigma_{\textrm{out}}(t) + \sigma_{\textrm{in}}(t)\right] - \sigma_{\textrm{out}}(t) - \sigma_{\textrm{in}}(t).
    \end{align}
    Multiplying the hyperbolic sine by the exponential prefactor in Eq.~\eqref{eq:Gink_inter}, we get
    \begin{align}\label{eq:exp_sinh}
        e^{-2\delta S} e^{-\sum_{k^\p} 2 f_{k^\p}^2 n_B(\omega_{k^\p})} \sinh\left[\sigma_{\textrm{out}}(t) + \sigma_{\textrm{in}}(t)\right] =& \nonumber \\
        &\hspace{-5cm}+\frac{e^{-2\delta S_2}}{2} \exp\left(-\sum_{k^\p}\left[f_{k^\p}^2\left\{\left[1+n_B(\omega_{k^\p})\right]\left(1 - e^{-\ii\omega_{k^\p}t}\right) + n_B(\omega_{k^\p})\left(1 - e^{\ii\omega_{k^\p}t}\right)\right\} + \tilde{f}_{k^\p}^2 - f_{k^\p}^2\right]\right) \nonumber \\
        &\hspace{-5cm}-\frac{e^{-2\delta S_2}}{2} \exp\left(-\sum_{k^\p}\left[f_{k^\p}^2\left\{\left[1+n_B(\omega_{k^\p})\right]\left(1 + e^{-\ii\omega_{k^\p}t}\right) + n_B(\omega_{k^\p})\left(1 + e^{\ii\omega_{k^\p}t}\right)\right\} + \tilde{f}_{k^\p}^2 - f_{k^\p}^2\right]\right).
    \end{align}
    
    \end{widetext}
    At low frequencies (with respect to the resonance frequency $\omega_0$), one has $f_k^2 \sim 2\Delta /(z\omega_k)$. Assuming open circuit boundary conditions at the far end of the array (that is, opposite from the transmon), the first mode is located at $\omega_k = \Delta/2$, such that the sum over the modes diverges logarithmically in $\Delta$, $\sum_{k^\p < \omega_c/v} f_{k^\p}^2 \sim (2/z)\log(\Delta /(2\omega_c))$, where $\omega_c < \omega_0$ is some cutoff frequency for the low-energy modes of the system. The exponential in the second line of Eq.~\eqref{eq:exp_sinh} vanishes in the thermodynamic limit as a power law in the system size, $\Delta^{2/z}\sim N^{-2/z}$. The first exponential remains finite due to the $1 - e^{\pm \ii\omega_{k^\p} t}$ factors (note that $\tilde{f}_k^2 \sim 2\Delta /(z\omega_k)$ at low frequencies as well, so that $\tilde{f}_k^2 - f_k^2 \rightarrow 0$ at $\omega_k\rightarrow 0$). The linear terms in Eq.~\eqref{eq:sigma_sum} vanish similarly when multiplied by $e^{-2 \delta S}$. We thus recover the integral expression for $\Gamma_k^{\textrm{in}}$ in Eq.~\eqref{eq:Gink}.
    
    \begin{figure*}[t!]
    	\centering
    	\includegraphics[width=0.98\textwidth]{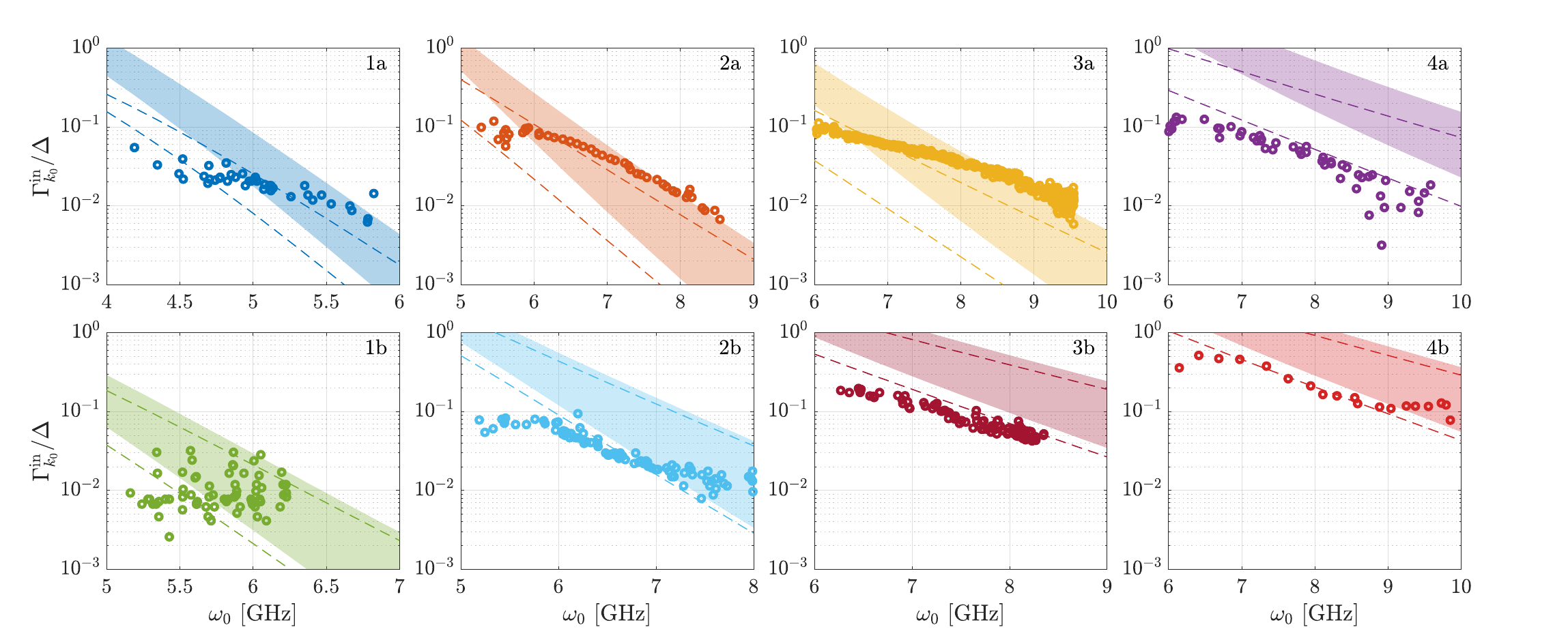}
    	\caption{Comparison of the on-resonance decay rate, calculated at a temperature $T=50~\textrm{mK}$, with the experimental data. See the caption of Fig.~\ref{fig:exp_comp} (calculated at $T=40~\textrm{mK}$) for details.}
    	\label{fig:exp_comp_high_T}
    \end{figure*}

    \subsection{Numerical evaluation of Eq.~\eqref{eq:Gink}}
    Finally, let us refer to an important detail in the numerical evaluation of the inelastic decay rate of near-resonance photons. While Eq.~\eqref{eq:Gink} provides a compact integral expression for $\Gamma_k^{\textrm{in}}$, its numerical evaluation could be challenging due to the time-oscillating exponentials in the sum over the modes. Indeed, the near-resonance part of the sum over $k^\p$ in Eq.~\eqref{eq:Gink} leads to oscillations of the integrand spanning many orders of magnitude, leading to ill-behaved time integration. Here we describe how one may overcome this obstacle. We begin by considering the zero-temperature case, then discuss the case $T \ll \omega_k$. At $T=0$, we have
        \begin{align}\label{eq:ImPiR_T=0}
            \Im\Pi_R(\omega_k;T=0) =& -\lambda_0^2 e^{-2\delta S} \nonumber \\
            &\hspace{-2cm}\times\Im \left\{\int_0^{\infty}\id t\sin(\omega_k t) e^{\sum_{k^\p}f_{k^\p}^2e^{-\ii\omega_{k^\p}t}}\right\}.
        \end{align}
        Let us introduce a frequency $\omega_c$ satisfying $\omega_k/2<\omega_c<\omega_k$, and split the sum over $k^\p$ to modes below and above $\omega_c$. Expanding the exponential of the $k^{\p}>\omega_c/v$ part leads to
        \begin{widetext}
        \begin{align}\label{eq:Gink_T=0_manipulation}
            \Im\Pi_R(\omega_k;T=0) =& -\lambda_0^2 e^{-2\delta S} \Im \Bigg\{\int_0^{\infty}\id t\sin(\omega_k t) e^{\sum_{k^\p < \omega_c/v}f_{k^\p}^2e^{-\ii\omega_{k^\p}t}}\nonumber \\
            &\hspace{2cm}\times\left(1 + \sum_{k^\p>\omega_c/v}f_{k^\p}^2 e^{-\ii\omega_{k^\p} t} + \frac{1}{2}\sum_{k^\p,k^{\p\p}>\omega_c/v}f_{k^\p}^2f_{k^{\p\p}}^2 e^{-\ii(\omega_{k^\p} +\omega_{k^{\p\p}})t}+\ldots\right)\Bigg\}.
        \end{align}
     Now, using $2\pi\delta(\omega) = \Re\left\{\int_0^{\infty} \id t e^{\ii\omega t}\right\}$, we find that the terms of second order and higher in the Taylor expansion of the $k^\p>\omega_c/v$ exponential do not contribute to the imaginary part of the self-energy. This is physically clear --- an incoming photon at frequency $\omega_k$ cannot produce more than a single outgoing photon at a frequency $\omega_{k^\p} > \omega_k/2$. This leads to a relation exactly equivalent to Eq.~\eqref{eq:ImPiR_T=0},
     \begin{equation}
            \Im\Pi_R(\omega_k;T=0) = \frac{\lambda_0^2}{2} e^{-2\delta S} \Re \left\{\int_0^{\infty}\id t e^{\ii\omega_k t} e^{\sum_{k^\p < \omega_c/v}f_{k^\p}^2e^{-\ii\omega_{k^\p}t}}\times\left(1 + \sum_{k^\p>\omega_c/v}f_{k^\p}^2 e^{-\ii\omega_{k^\p} t}\right)\right\},
        \end{equation}
        which we reorganize as
        \begin{equation}
            \Im\Pi_R(\omega_k;T=0) = \frac{\lambda_0^2}{2} e^{-2\delta S} e^{\sum_{k^\p < \omega_c/v}f_{k^\p}^2} \Re \left\{\int_0^{\infty}\id t e^{\ii\omega_k t} e^{-\sum_{k^\p < \omega_c/v}f_{k^\p}^2\left(1-e^{-\ii\omega_{k^\p}t}\right)}\times\left(1 + \sum_{k^\p>\omega_c/v}f_{k^\p}^2 e^{-\ii\omega_{k^\p} t}\right)\right\}.
        \end{equation}
For a near-resonance incoming photon, $\omega_k\sim\omega_0$, this form is much more favorable for numerical evaluation, as the strong oscillations of the integrand are absent from this expression.

    Generalization to finite temperature, where $T \ll \omega_k$, is straightforward, by keeping a small but finite number of terms in the expansion of the $k^\p >\omega_c/v$ part; namely,
        \begin{align}\label{eq:Gink_finite_T_manipulation}
            \Im\Pi_R(\omega_k;T\ll\omega_k) \approx& -\lambda_0^2 e^{-2\delta S}e^{\sum_{k^\p < \omega_c/v}f_{k^\p}^2} \Im \Bigg\{\int_0^{\infty}\id t\sin(\omega_k t) \sum_{m=0}^M\frac{1}{m!}\left(\sum_{k^\p>\omega_c/v}f_{k^\p}^2 e^{-\ii\omega_{k^\p} t}\right)^m \nonumber \\
            &\hspace{2cm}\times \exp\left(-\sum_{k^\p < \omega_c/v}f_{k^\p}^2\left(1 - e^{-\ii\omega_{k^\p}t}\right) - \sum_{k^\p}2f_{k^\p}^2 n_B(\omega_{k^\p})\left(1 - \cos(\omega_{k^\p}t)\right)\right)\Bigg\}.
        \end{align}
        Note that in contrast to Eq.~\eqref{eq:Gink_T=0_manipulation}, Eq.~\eqref{eq:Gink_finite_T_manipulation} is an approximation of the imaginary part of Eq.~\eqref{eq:Gink}. However, one may verify that the contribution of the terms from the expansion of the $k^\p >\omega_c/v$ part decay rapidly, and choose a large enough number of terms ($M = 5$ is enough for realistic parameters) to achieve the desired numerical precision.
        \end{widetext}
        
        \section{Results assuming higher temperature}\label{app:temp}
        In producing Fig.~\ref{fig:exp_comp}, we assumed a uniform temperature of $T = 40~\textrm{mK}$ for all devices. However, the temperature cannot be measured directly; as mentioned above, one may only estimate a typical temperature range from qubit experiments on similar setups. Furthermore, the temperature may deviate between experiments. In Fig.~\ref{fig:exp_comp_high_T}, we show a comparison of the experimental measurements with the theory, calculated at $T = 50~\textrm{mK}$, which is also within the range of typical temperatures. This higher temperature provides better fits for the low impedance devices (other than 4a, which has the smallest $\omega_0/\Gamma_0$), and slightly worsens the fit for the high impedance devices.
        

        \bibliography{num_inst_bib.bib}

\begin{thebibliography}{44}%
\makeatletter
\providecommand \@ifxundefined [1]{%
 \@ifx{#1\undefined}
}%
\providecommand \@ifnum [1]{%
 \ifnum #1\expandafter \@firstoftwo
 \else \expandafter \@secondoftwo
 \fi
}%
\providecommand \@ifx [1]{%
 \ifx #1\expandafter \@firstoftwo
 \else \expandafter \@secondoftwo
 \fi
}%
\providecommand \natexlab [1]{#1}%
\providecommand \enquote  [1]{``#1''}%
\providecommand \bibnamefont  [1]{#1}%
\providecommand \bibfnamefont [1]{#1}%
\providecommand \citenamefont [1]{#1}%
\providecommand \href@noop [0]{\@secondoftwo}%
\providecommand \href [0]{\begingroup \@sanitize@url \@href}%
\providecommand \@href[1]{\@@startlink{#1}\@@href}%
\providecommand \@@href[1]{\endgroup#1\@@endlink}%
\providecommand \@sanitize@url [0]{\catcode `\\12\catcode `\$12\catcode
  `\&12\catcode `\#12\catcode `\^12\catcode `\_12\catcode `\%12\relax}%
\providecommand \@@startlink[1]{}%
\providecommand \@@endlink[0]{}%
\providecommand \url  [0]{\begingroup\@sanitize@url \@url }%
\providecommand \@url [1]{\endgroup\@href {#1}{\urlprefix }}%
\providecommand \urlprefix  [0]{URL }%
\providecommand \Eprint [0]{\href }%
\providecommand \doibase [0]{https://doi.org/}%
\providecommand \selectlanguage [0]{\@gobble}%
\providecommand \bibinfo  [0]{\@secondoftwo}%
\providecommand \bibfield  [0]{\@secondoftwo}%
\providecommand \translation [1]{[#1]}%
\providecommand \BibitemOpen [0]{}%
\providecommand \bibitemStop [0]{}%
\providecommand \bibitemNoStop [0]{.\EOS\space}%
\providecommand \EOS [0]{\spacefactor3000\relax}%
\providecommand \BibitemShut  [1]{\csname bibitem#1\endcsname}%
\let\auto@bib@innerbib\@empty
\bibitem [{\citenamefont {Zamolodchikov}\ and\ \citenamefont
  {Zamolodchikov}(1979)}]{zamolodchikov_factorized_1979}%
  \BibitemOpen
  \bibfield  {author} {\bibinfo {author} {\bibfnamefont {A.~B.}\ \bibnamefont
  {Zamolodchikov}}\ and\ \bibinfo {author} {\bibfnamefont {A.~B.}\ \bibnamefont
  {Zamolodchikov}},\ }\bibfield  {title} {\bibinfo {title} {Factorized
  \textit{{S}}-matrices in two dimensions as the exact solutions of certain
  relativistic quantum field theory models},\ }\href
  {https://doi.org/10.1016/0003-4916(79)90391-9} {\bibfield  {journal}
  {\bibinfo  {journal} {Annals of Physics}\ }\textbf {\bibinfo {volume}
  {120}},\ \bibinfo {pages} {253} (\bibinfo {year} {1979})}\BibitemShut
  {NoStop}%
\bibitem [{\citenamefont {Leggett}\ \emph {et~al.}(1987)\citenamefont
  {Leggett}, \citenamefont {Chakravarty}, \citenamefont {Dorsey}, \citenamefont
  {Fisher}, \citenamefont {Garg},\ and\ \citenamefont
  {Zwerger}}]{leggett_dynamics_1987}%
  \BibitemOpen
  \bibfield  {author} {\bibinfo {author} {\bibfnamefont {A.~J.}\ \bibnamefont
  {Leggett}}, \bibinfo {author} {\bibfnamefont {S.}~\bibnamefont
  {Chakravarty}}, \bibinfo {author} {\bibfnamefont {A.~T.}\ \bibnamefont
  {Dorsey}}, \bibinfo {author} {\bibfnamefont {M.~P.~A.}\ \bibnamefont
  {Fisher}}, \bibinfo {author} {\bibfnamefont {A.}~\bibnamefont {Garg}},\ and\
  \bibinfo {author} {\bibfnamefont {W.}~\bibnamefont {Zwerger}},\ }\bibfield
  {title} {\bibinfo {title} {Dynamics of the dissipative two-state system},\
  }\href {https://doi.org/10.1103/RevModPhys.59.1} {\bibfield  {journal}
  {\bibinfo  {journal} {Reviews of Modern Physics}\ }\textbf {\bibinfo {volume}
  {59}},\ \bibinfo {pages} {1} (\bibinfo {year} {1987})}\BibitemShut {NoStop}%
\bibitem [{\citenamefont {Gross}\ \emph {et~al.}(1981)\citenamefont {Gross},
  \citenamefont {Pisarski},\ and\ \citenamefont {Yaffe}}]{gross_qcd_1981}%
  \BibitemOpen
  \bibfield  {author} {\bibinfo {author} {\bibfnamefont {D.~J.}\ \bibnamefont
  {Gross}}, \bibinfo {author} {\bibfnamefont {R.~D.}\ \bibnamefont
  {Pisarski}},\ and\ \bibinfo {author} {\bibfnamefont {L.~G.}\ \bibnamefont
  {Yaffe}},\ }\bibfield  {title} {\bibinfo {title} {{QCD} and instantons at
  finite temperature},\ }\href {https://doi.org/10.1103/RevModPhys.53.43}
  {\bibfield  {journal} {\bibinfo  {journal} {Reviews of Modern Physics}\
  }\textbf {\bibinfo {volume} {53}},\ \bibinfo {pages} {43} (\bibinfo {year}
  {1981})}\BibitemShut {NoStop}%
\bibitem [{\citenamefont {Vandoren}\ and\ \citenamefont {van
  Nieuwenhuizen}(2008)}]{vandoren_lectures_2008}%
  \BibitemOpen
  \bibfield  {author} {\bibinfo {author} {\bibfnamefont {S.}~\bibnamefont
  {Vandoren}}\ and\ \bibinfo {author} {\bibfnamefont {P.}~\bibnamefont {van
  Nieuwenhuizen}},\ }\href {https://doi.org/10.48550/arXiv.0802.1862} {\bibinfo
  {title} {Lectures on instantons}} (\bibinfo {year} {2008}),\ \bibinfo {note}
  {arXiv:0802.1862 [hep-th]}\BibitemShut {NoStop}%
\bibitem [{\citenamefont {Schön}\ and\ \citenamefont
  {Zaikin}(1990)}]{schon_quantum_1990}%
  \BibitemOpen
  \bibfield  {author} {\bibinfo {author} {\bibfnamefont {G.}~\bibnamefont
  {Schön}}\ and\ \bibinfo {author} {\bibfnamefont {A.~D.}\ \bibnamefont
  {Zaikin}},\ }\bibfield  {title} {\bibinfo {title} {Quantum coherent effects,
  phase transitions, and the dissipative dynamics of ultra small tunnel
  junctions},\ }\href {https://doi.org/10.1016/0370-1573(90)90156-V} {\bibfield
   {journal} {\bibinfo  {journal} {Physics Reports}\ }\textbf {\bibinfo
  {volume} {198}},\ \bibinfo {pages} {237} (\bibinfo {year}
  {1990})}\BibitemShut {NoStop}%
\bibitem [{\citenamefont {Fazio}\ and\ \citenamefont {van~der
  Zant}(2001)}]{fazio_quantum_2001}%
  \BibitemOpen
  \bibfield  {author} {\bibinfo {author} {\bibfnamefont {R.}~\bibnamefont
  {Fazio}}\ and\ \bibinfo {author} {\bibfnamefont {H.}~\bibnamefont {van~der
  Zant}},\ }\bibfield  {title} {\bibinfo {title} {Quantum phase transitions and
  vortex dynamics in superconducting networks},\ }\href
  {https://doi.org/10.1016/S0370-1573(01)00022-9} {\bibfield  {journal}
  {\bibinfo  {journal} {Physics Reports}\ }\textbf {\bibinfo {volume} {355}},\
  \bibinfo {pages} {235} (\bibinfo {year} {2001})}\BibitemShut {NoStop}%
\bibitem [{\citenamefont {Schmid}(1983)}]{schmid_diffusion_1983}%
  \BibitemOpen
  \bibfield  {author} {\bibinfo {author} {\bibfnamefont {A.}~\bibnamefont
  {Schmid}},\ }\bibfield  {title} {\bibinfo {title} {Diffusion and
  {Localization} in a {Dissipative} {Quantum} {System}},\ }\href
  {https://doi.org/10.1103/PhysRevLett.51.1506} {\bibfield  {journal} {\bibinfo
   {journal} {Physical Review Letters}\ }\textbf {\bibinfo {volume} {51}},\
  \bibinfo {pages} {1506} (\bibinfo {year} {1983})}\BibitemShut {NoStop}%
\bibitem [{\citenamefont {Bulgadaev}(1984)}]{bulgadaev_phase_1984}%
  \BibitemOpen
  \bibfield  {author} {\bibinfo {author} {\bibfnamefont {S.}~\bibnamefont
  {Bulgadaev}},\ }\bibfield  {title} {\bibinfo {title} {Phase diagram of a
  dissipative quantum system},\ }\href@noop {} {\bibfield  {journal} {\bibinfo
  {journal} {ZhETF Pisma Redaktsiiu}\ } (\bibinfo {year} {1984})}\BibitemShut
  {NoStop}%
\bibitem [{\citenamefont {Rastelli}\ \emph {et~al.}(2013)\citenamefont
  {Rastelli}, \citenamefont {Pop},\ and\ \citenamefont
  {Hekking}}]{rastelli_quantum_2013}%
  \BibitemOpen
  \bibfield  {author} {\bibinfo {author} {\bibfnamefont {G.}~\bibnamefont
  {Rastelli}}, \bibinfo {author} {\bibfnamefont {I.~M.}\ \bibnamefont {Pop}},\
  and\ \bibinfo {author} {\bibfnamefont {F.~W.~J.}\ \bibnamefont {Hekking}},\
  }\bibfield  {title} {\bibinfo {title} {Quantum phase slips in {Josephson}
  junction rings},\ }\href {https://doi.org/10.1103/PhysRevB.87.174513}
  {\bibfield  {journal} {\bibinfo  {journal} {Physical Review B}\ }\textbf
  {\bibinfo {volume} {87}},\ \bibinfo {pages} {174513} (\bibinfo {year}
  {2013})}\BibitemShut {NoStop}%
\bibitem [{\citenamefont {Bard}\ \emph {et~al.}(2017)\citenamefont {Bard},
  \citenamefont {Protopopov}, \citenamefont {Gornyi}, \citenamefont
  {Shnirman},\ and\ \citenamefont
  {Mirlin}}]{bard_superconductor-insulator_2017}%
  \BibitemOpen
  \bibfield  {author} {\bibinfo {author} {\bibfnamefont {M.}~\bibnamefont
  {Bard}}, \bibinfo {author} {\bibfnamefont {I.~V.}\ \bibnamefont
  {Protopopov}}, \bibinfo {author} {\bibfnamefont {I.~V.}\ \bibnamefont
  {Gornyi}}, \bibinfo {author} {\bibfnamefont {A.}~\bibnamefont {Shnirman}},\
  and\ \bibinfo {author} {\bibfnamefont {A.~D.}\ \bibnamefont {Mirlin}},\
  }\bibfield  {title} {\bibinfo {title} {Superconductor-insulator transition in
  disordered {Josephson}-junction chains},\ }\href
  {https://doi.org/10.1103/PhysRevB.96.064514} {\bibfield  {journal} {\bibinfo
  {journal} {Physical Review B}\ }\textbf {\bibinfo {volume} {96}},\ \bibinfo
  {pages} {064514} (\bibinfo {year} {2017})}\BibitemShut {NoStop}%
\bibitem [{\citenamefont {Kästner}\ and\ \citenamefont
  {Kozuch}(2020)}]{kastner_tunnelling_2020}%
  \BibitemOpen
  \bibfield  {author} {\bibinfo {author} {\bibfnamefont {J.}~\bibnamefont
  {Kästner}}\ and\ \bibinfo {author} {\bibfnamefont {S.}~\bibnamefont
  {Kozuch}},\ }\href {https://doi.org/10.1039/9781839160370} {\emph {\bibinfo
  {title} {Tunnelling in {Molecules}: {Nuclear} {Quantum} {Effects} from {Bio}
  to {Physical} {Chemistry}}}}\ (\bibinfo  {publisher} {The Royal Society of
  Chemistry},\ \bibinfo {year} {2020})\BibitemShut {NoStop}%
\bibitem [{\citenamefont {Koch}\ \emph {et~al.}(2007)\citenamefont {Koch},
  \citenamefont {Yu}, \citenamefont {Gambetta}, \citenamefont {Houck},
  \citenamefont {Schuster}, \citenamefont {Majer}, \citenamefont {Blais},
  \citenamefont {Devoret}, \citenamefont {Girvin},\ and\ \citenamefont
  {Schoelkopf}}]{koch_charge-insensitive_2007}%
  \BibitemOpen
  \bibfield  {author} {\bibinfo {author} {\bibfnamefont {J.}~\bibnamefont
  {Koch}}, \bibinfo {author} {\bibfnamefont {T.~M.}\ \bibnamefont {Yu}},
  \bibinfo {author} {\bibfnamefont {J.}~\bibnamefont {Gambetta}}, \bibinfo
  {author} {\bibfnamefont {A.~A.}\ \bibnamefont {Houck}}, \bibinfo {author}
  {\bibfnamefont {D.~I.}\ \bibnamefont {Schuster}}, \bibinfo {author}
  {\bibfnamefont {J.}~\bibnamefont {Majer}}, \bibinfo {author} {\bibfnamefont
  {A.}~\bibnamefont {Blais}}, \bibinfo {author} {\bibfnamefont {M.~H.}\
  \bibnamefont {Devoret}}, \bibinfo {author} {\bibfnamefont {S.~M.}\
  \bibnamefont {Girvin}},\ and\ \bibinfo {author} {\bibfnamefont {R.~J.}\
  \bibnamefont {Schoelkopf}},\ }\bibfield  {title} {\bibinfo {title}
  {Charge-insensitive qubit design derived from the {Cooper} pair box},\ }\href
  {https://doi.org/10.1103/PhysRevA.76.042319} {\bibfield  {journal} {\bibinfo
  {journal} {Physical Review A}\ }\textbf {\bibinfo {volume} {76}},\ \bibinfo
  {pages} {042319} (\bibinfo {year} {2007})}\BibitemShut {NoStop}%
\bibitem [{\citenamefont {Goldstein}\ \emph {et~al.}(2013)\citenamefont
  {Goldstein}, \citenamefont {Devoret}, \citenamefont {Houzet},\ and\
  \citenamefont {Glazman}}]{goldstein_inelastic_2013}%
  \BibitemOpen
  \bibfield  {author} {\bibinfo {author} {\bibfnamefont {M.}~\bibnamefont
  {Goldstein}}, \bibinfo {author} {\bibfnamefont {M.~H.}\ \bibnamefont
  {Devoret}}, \bibinfo {author} {\bibfnamefont {M.}~\bibnamefont {Houzet}},\
  and\ \bibinfo {author} {\bibfnamefont {L.~I.}\ \bibnamefont {Glazman}},\
  }\bibfield  {title} {\bibinfo {title} {Inelastic {Microwave} {Photon}
  {Scattering} off a {Quantum} {Impurity} in a {Josephson}-{Junction}
  {Array}},\ }\href {https://doi.org/10.1103/PhysRevLett.110.017002} {\bibfield
   {journal} {\bibinfo  {journal} {Physical Review Letters}\ }\textbf {\bibinfo
  {volume} {110}},\ \bibinfo {pages} {017002} (\bibinfo {year}
  {2013})}\BibitemShut {NoStop}%
\bibitem [{\citenamefont {Kuzmin}\ \emph
  {et~al.}(2019{\natexlab{a}})\citenamefont {Kuzmin}, \citenamefont {Mehta},
  \citenamefont {Grabon}, \citenamefont {Mencia},\ and\ \citenamefont
  {Manucharyan}}]{kuzmin_superstrong_2019}%
  \BibitemOpen
  \bibfield  {author} {\bibinfo {author} {\bibfnamefont {R.}~\bibnamefont
  {Kuzmin}}, \bibinfo {author} {\bibfnamefont {N.}~\bibnamefont {Mehta}},
  \bibinfo {author} {\bibfnamefont {N.}~\bibnamefont {Grabon}}, \bibinfo
  {author} {\bibfnamefont {R.}~\bibnamefont {Mencia}},\ and\ \bibinfo {author}
  {\bibfnamefont {V.~E.}\ \bibnamefont {Manucharyan}},\ }\bibfield  {title}
  {\bibinfo {title} {Superstrong coupling in circuit quantum electrodynamics},\
  }\href {https://doi.org/10.1038/s41534-019-0134-2} {\bibfield  {journal}
  {\bibinfo  {journal} {npj Quantum Information}\ }\textbf {\bibinfo {volume}
  {5}},\ \bibinfo {pages} {1} (\bibinfo {year}
  {2019}{\natexlab{a}})}\BibitemShut {NoStop}%
\bibitem [{\citenamefont {Puertas~Martínez}\ \emph {et~al.}(2019)\citenamefont
  {Puertas~Martínez}, \citenamefont {Léger}, \citenamefont {Gheeraert},
  \citenamefont {Dassonneville}, \citenamefont {Planat}, \citenamefont
  {Foroughi}, \citenamefont {Krupko}, \citenamefont {Buisson}, \citenamefont
  {Naud}, \citenamefont {Hasch-Guichard}, \citenamefont {Florens},
  \citenamefont {Snyman},\ and\ \citenamefont
  {Roch}}]{puertas_martinez_tunable_2019}%
  \BibitemOpen
  \bibfield  {author} {\bibinfo {author} {\bibfnamefont {J.}~\bibnamefont
  {Puertas~Martínez}}, \bibinfo {author} {\bibfnamefont {S.}~\bibnamefont
  {Léger}}, \bibinfo {author} {\bibfnamefont {N.}~\bibnamefont {Gheeraert}},
  \bibinfo {author} {\bibfnamefont {R.}~\bibnamefont {Dassonneville}}, \bibinfo
  {author} {\bibfnamefont {L.}~\bibnamefont {Planat}}, \bibinfo {author}
  {\bibfnamefont {F.}~\bibnamefont {Foroughi}}, \bibinfo {author}
  {\bibfnamefont {Y.}~\bibnamefont {Krupko}}, \bibinfo {author} {\bibfnamefont
  {O.}~\bibnamefont {Buisson}}, \bibinfo {author} {\bibfnamefont
  {C.}~\bibnamefont {Naud}}, \bibinfo {author} {\bibfnamefont {W.}~\bibnamefont
  {Hasch-Guichard}}, \bibinfo {author} {\bibfnamefont {S.}~\bibnamefont
  {Florens}}, \bibinfo {author} {\bibfnamefont {I.}~\bibnamefont {Snyman}},\
  and\ \bibinfo {author} {\bibfnamefont {N.}~\bibnamefont {Roch}},\ }\bibfield
  {title} {\bibinfo {title} {A tunable {Josephson} platform to explore
  many-body quantum optics in circuit-{QED}},\ }\href
  {https://doi.org/10.1038/s41534-018-0104-0} {\bibfield  {journal} {\bibinfo
  {journal} {npj Quantum Information}\ }\textbf {\bibinfo {volume} {5}},\
  \bibinfo {pages} {1} (\bibinfo {year} {2019})}\BibitemShut {NoStop}%
\bibitem [{\citenamefont {Bard}\ \emph {et~al.}(2018)\citenamefont {Bard},
  \citenamefont {Protopopov},\ and\ \citenamefont {Mirlin}}]{bard_decay_2018}%
  \BibitemOpen
  \bibfield  {author} {\bibinfo {author} {\bibfnamefont {M.}~\bibnamefont
  {Bard}}, \bibinfo {author} {\bibfnamefont {I.~V.}\ \bibnamefont
  {Protopopov}},\ and\ \bibinfo {author} {\bibfnamefont {A.~D.}\ \bibnamefont
  {Mirlin}},\ }\bibfield  {title} {\bibinfo {title} {Decay of plasmonic waves
  in {Josephson} junction chains},\ }\href
  {https://doi.org/10.1103/PhysRevB.98.224513} {\bibfield  {journal} {\bibinfo
  {journal} {Physical Review B}\ }\textbf {\bibinfo {volume} {98}},\ \bibinfo
  {pages} {224513} (\bibinfo {year} {2018})}\BibitemShut {NoStop}%
\bibitem [{\citenamefont {Wu}\ and\ \citenamefont
  {Sau}(2019)}]{wu_theory_2019}%
  \BibitemOpen
  \bibfield  {author} {\bibinfo {author} {\bibfnamefont {H.-K.}\ \bibnamefont
  {Wu}}\ and\ \bibinfo {author} {\bibfnamefont {J.~D.}\ \bibnamefont {Sau}},\
  }\bibfield  {title} {\bibinfo {title} {Theory of coherent phase modes in
  insulating {Josephson} junction chains},\ }\href
  {https://doi.org/10.1103/PhysRevB.99.214509} {\bibfield  {journal} {\bibinfo
  {journal} {Physical Review B}\ }\textbf {\bibinfo {volume} {99}},\ \bibinfo
  {pages} {214509} (\bibinfo {year} {2019})}\BibitemShut {NoStop}%
\bibitem [{\citenamefont {Kuzmin}\ \emph
  {et~al.}(2019{\natexlab{b}})\citenamefont {Kuzmin}, \citenamefont {Mencia},
  \citenamefont {Grabon}, \citenamefont {Mehta}, \citenamefont {Lin},\ and\
  \citenamefont {Manucharyan}}]{kuzmin_quantum_2019}%
  \BibitemOpen
  \bibfield  {author} {\bibinfo {author} {\bibfnamefont {R.}~\bibnamefont
  {Kuzmin}}, \bibinfo {author} {\bibfnamefont {R.}~\bibnamefont {Mencia}},
  \bibinfo {author} {\bibfnamefont {N.}~\bibnamefont {Grabon}}, \bibinfo
  {author} {\bibfnamefont {N.}~\bibnamefont {Mehta}}, \bibinfo {author}
  {\bibfnamefont {Y.-H.}\ \bibnamefont {Lin}},\ and\ \bibinfo {author}
  {\bibfnamefont {V.~E.}\ \bibnamefont {Manucharyan}},\ }\bibfield  {title}
  {\bibinfo {title} {Quantum electrodynamics of a superconductor–insulator
  phase transition},\ }\href {https://doi.org/10.1038/s41567-019-0553-1}
  {\bibfield  {journal} {\bibinfo  {journal} {Nature Physics}\ }\textbf
  {\bibinfo {volume} {15}},\ \bibinfo {pages} {930} (\bibinfo {year}
  {2019}{\natexlab{b}})}\BibitemShut {NoStop}%
\bibitem [{\citenamefont {Houzet}\ and\ \citenamefont
  {Glazman}(2020)}]{houzet_critical_2020}%
  \BibitemOpen
  \bibfield  {author} {\bibinfo {author} {\bibfnamefont {M.}~\bibnamefont
  {Houzet}}\ and\ \bibinfo {author} {\bibfnamefont {L.}~\bibnamefont
  {Glazman}},\ }\bibfield  {title} {\bibinfo {title} {Critical {Fluorescence}
  of a {Transmon} at the {Schmid} {Transition}},\ }\href
  {https://doi.org/10.1103/PhysRevLett.125.267701} {\bibfield  {journal}
  {\bibinfo  {journal} {Physical Review Letters}\ }\textbf {\bibinfo {volume}
  {125}},\ \bibinfo {pages} {267701} (\bibinfo {year} {2020})}\BibitemShut
  {NoStop}%
\bibitem [{\citenamefont {Burshtein}\ \emph {et~al.}(2021)\citenamefont
  {Burshtein}, \citenamefont {Kuzmin}, \citenamefont {Manucharyan},\ and\
  \citenamefont {Goldstein}}]{burshtein_photon-instanton_2021}%
  \BibitemOpen
  \bibfield  {author} {\bibinfo {author} {\bibfnamefont {A.}~\bibnamefont
  {Burshtein}}, \bibinfo {author} {\bibfnamefont {R.}~\bibnamefont {Kuzmin}},
  \bibinfo {author} {\bibfnamefont {V.~E.}\ \bibnamefont {Manucharyan}},\ and\
  \bibinfo {author} {\bibfnamefont {M.}~\bibnamefont {Goldstein}},\ }\bibfield
  {title} {\bibinfo {title} {Photon-{Instanton} {Collider} {Implemented} by a
  {Superconducting} {Circuit}},\ }\href
  {https://doi.org/10.1103/PhysRevLett.126.137701} {\bibfield  {journal}
  {\bibinfo  {journal} {Physical Review Letters}\ }\textbf {\bibinfo {volume}
  {126}},\ \bibinfo {pages} {137701} (\bibinfo {year} {2021})}\BibitemShut
  {NoStop}%
\bibitem [{\citenamefont {Kuzmin}\ \emph {et~al.}(2021)\citenamefont {Kuzmin},
  \citenamefont {Grabon}, \citenamefont {Mehta}, \citenamefont {Burshtein},
  \citenamefont {Goldstein}, \citenamefont {Houzet}, \citenamefont {Glazman},\
  and\ \citenamefont {Manucharyan}}]{kuzmin_inelastic_2021}%
  \BibitemOpen
  \bibfield  {author} {\bibinfo {author} {\bibfnamefont {R.}~\bibnamefont
  {Kuzmin}}, \bibinfo {author} {\bibfnamefont {N.}~\bibnamefont {Grabon}},
  \bibinfo {author} {\bibfnamefont {N.}~\bibnamefont {Mehta}}, \bibinfo
  {author} {\bibfnamefont {A.}~\bibnamefont {Burshtein}}, \bibinfo {author}
  {\bibfnamefont {M.}~\bibnamefont {Goldstein}}, \bibinfo {author}
  {\bibfnamefont {M.}~\bibnamefont {Houzet}}, \bibinfo {author} {\bibfnamefont
  {L.}~\bibnamefont {Glazman}},\ and\ \bibinfo {author} {\bibfnamefont
  {V.}~\bibnamefont {Manucharyan}},\ }\bibfield  {title} {\bibinfo {title}
  {Inelastic {Scattering} of a {Photon} by a {Quantum} {Phase} {Slip}},\ }\href
  {https://doi.org/10.1103/PhysRevLett.126.197701} {\bibfield  {journal}
  {\bibinfo  {journal} {Physical Review Letters}\ }\textbf {\bibinfo {volume}
  {126}},\ \bibinfo {pages} {197701} (\bibinfo {year} {2021})}\BibitemShut
  {NoStop}%
\bibitem [{\citenamefont {Mehta}\ \emph {et~al.}(2023)\citenamefont {Mehta},
  \citenamefont {Kuzmin}, \citenamefont {Ciuti},\ and\ \citenamefont
  {Manucharyan}}]{mehta_down-conversion_2023}%
  \BibitemOpen
  \bibfield  {author} {\bibinfo {author} {\bibfnamefont {N.}~\bibnamefont
  {Mehta}}, \bibinfo {author} {\bibfnamefont {R.}~\bibnamefont {Kuzmin}},
  \bibinfo {author} {\bibfnamefont {C.}~\bibnamefont {Ciuti}},\ and\ \bibinfo
  {author} {\bibfnamefont {V.~E.}\ \bibnamefont {Manucharyan}},\ }\bibfield
  {title} {\bibinfo {title} {Down-conversion of a single photon as a probe of
  many-body localization},\ }\href {https://doi.org/10.1038/s41586-022-05615-y}
  {\bibfield  {journal} {\bibinfo  {journal} {Nature}\ }\textbf {\bibinfo
  {volume} {613}},\ \bibinfo {pages} {650} (\bibinfo {year}
  {2023})}\BibitemShut {NoStop}%
\bibitem [{\citenamefont {Mehta}\ \emph {et~al.}(2022)\citenamefont {Mehta},
  \citenamefont {Ciuti}, \citenamefont {Kuzmin},\ and\ \citenamefont
  {Manucharyan}}]{mehta_theory_2022}%
  \BibitemOpen
  \bibfield  {author} {\bibinfo {author} {\bibfnamefont {N.}~\bibnamefont
  {Mehta}}, \bibinfo {author} {\bibfnamefont {C.}~\bibnamefont {Ciuti}},
  \bibinfo {author} {\bibfnamefont {R.}~\bibnamefont {Kuzmin}},\ and\ \bibinfo
  {author} {\bibfnamefont {V.~E.}\ \bibnamefont {Manucharyan}},\ }\href
  {https://doi.org/10.48550/arXiv.2210.14681} {\bibinfo {title} {Theory of
  strong down-conversion in multi-mode cavity and circuit {QED}}} (\bibinfo
  {year} {2022}),\ \bibinfo {note} {arXiv:2210.14681}\BibitemShut {NoStop}%
\bibitem [{\citenamefont {Kuzmin}\ \emph {et~al.}(2025)\citenamefont {Kuzmin},
  \citenamefont {Mehta}, \citenamefont {Grabon}, \citenamefont {Mencia},
  \citenamefont {Burshtein}, \citenamefont {Goldstein},\ and\ \citenamefont
  {Manucharyan}}]{kuzmin_observation_2023}%
  \BibitemOpen
  \bibfield  {author} {\bibinfo {author} {\bibfnamefont {R.}~\bibnamefont
  {Kuzmin}}, \bibinfo {author} {\bibfnamefont {N.}~\bibnamefont {Mehta}},
  \bibinfo {author} {\bibfnamefont {N.}~\bibnamefont {Grabon}}, \bibinfo
  {author} {\bibfnamefont {R.~A.}\ \bibnamefont {Mencia}}, \bibinfo {author}
  {\bibfnamefont {A.}~\bibnamefont {Burshtein}}, \bibinfo {author}
  {\bibfnamefont {M.}~\bibnamefont {Goldstein}},\ and\ \bibinfo {author}
  {\bibfnamefont {V.~E.}\ \bibnamefont {Manucharyan}},\ }\bibfield  {title}
  {\bibinfo {title} {Observation of the schmid--bulgadaev dissipative quantum
  phase transition},\ }\href {https://doi.org/10.1038/s41567-024-02695-7}
  {\bibfield  {journal} {\bibinfo  {journal} {Nature Physics}\ }\textbf
  {\bibinfo {volume} {21}},\ \bibinfo {pages} {132} (\bibinfo {year}
  {2025})}\BibitemShut {NoStop}%
\bibitem [{\citenamefont {Burshtein}\ and\ \citenamefont
  {Goldstein}(2024)}]{burshtein_inelastic_2024}%
  \BibitemOpen
  \bibfield  {author} {\bibinfo {author} {\bibfnamefont {A.}~\bibnamefont
  {Burshtein}}\ and\ \bibinfo {author} {\bibfnamefont {M.}~\bibnamefont
  {Goldstein}},\ }\bibfield  {title} {\bibinfo {title} {Inelastic {Decay} from
  {Integrability}},\ }\href {https://doi.org/10.1103/PRXQuantum.5.020323}
  {\bibfield  {journal} {\bibinfo  {journal} {PRX Quantum}\ }\textbf {\bibinfo
  {volume} {5}},\ \bibinfo {pages} {020323} (\bibinfo {year}
  {2024})}\BibitemShut {NoStop}%
\bibitem [{\citenamefont {Houzet}\ \emph {et~al.}(2024)\citenamefont {Houzet},
  \citenamefont {Yamamoto},\ and\ \citenamefont
  {Glazman}}]{houzet_microwave_2024}%
  \BibitemOpen
  \bibfield  {author} {\bibinfo {author} {\bibfnamefont {M.}~\bibnamefont
  {Houzet}}, \bibinfo {author} {\bibfnamefont {T.}~\bibnamefont {Yamamoto}},\
  and\ \bibinfo {author} {\bibfnamefont {L.~I.}\ \bibnamefont {Glazman}},\
  }\bibfield  {title} {\bibinfo {title} {Microwave spectroscopy of the {Schmid}
  transition},\ }\href {https://doi.org/10.1103/PhysRevB.109.155431} {\bibfield
   {journal} {\bibinfo  {journal} {Physical Review B}\ }\textbf {\bibinfo
  {volume} {109}},\ \bibinfo {pages} {155431} (\bibinfo {year}
  {2024})}\BibitemShut {NoStop}%
\bibitem [{\citenamefont {Remez}\ \emph {et~al.}(2024)\citenamefont {Remez},
  \citenamefont {Kurilovich}, \citenamefont {Rieger},\ and\ \citenamefont
  {Glazman}}]{remez_bloch_2024}%
  \BibitemOpen
  \bibfield  {author} {\bibinfo {author} {\bibfnamefont {B.}~\bibnamefont
  {Remez}}, \bibinfo {author} {\bibfnamefont {V.~D.}\ \bibnamefont
  {Kurilovich}}, \bibinfo {author} {\bibfnamefont {M.}~\bibnamefont {Rieger}},\
  and\ \bibinfo {author} {\bibfnamefont {L.~I.}\ \bibnamefont {Glazman}},\
  }\bibfield  {title} {\bibinfo {title} {Bloch oscillations in a transmon
  embedded in a resonant electromagnetic environment},\ }\href
  {https://doi.org/10.1103/PhysRevB.110.054508} {\bibfield  {journal} {\bibinfo
   {journal} {Physical Review B}\ }\textbf {\bibinfo {volume} {110}},\ \bibinfo
  {pages} {054508} (\bibinfo {year} {2024})}\BibitemShut {NoStop}%
\bibitem [{\citenamefont {Burshtein}\ and\ \citenamefont
  {Goldstein}(2025)}]{burshtein_quantum_2024}%
  \BibitemOpen
  \bibfield  {author} {\bibinfo {author} {\bibfnamefont {A.}~\bibnamefont
  {Burshtein}}\ and\ \bibinfo {author} {\bibfnamefont {M.}~\bibnamefont
  {Goldstein}},\ }\bibfield  {title} {\bibinfo {title} {Quantum simulation of
  the microscopic to macroscopic crossover using superconducting quantum
  impurities},\ }\href {https://doi.org/10.1103/PhysRevB.111.174303} {\bibfield
   {journal} {\bibinfo  {journal} {Phys. Rev. B}\ }\textbf {\bibinfo {volume}
  {111}},\ \bibinfo {pages} {174303} (\bibinfo {year} {2025})}\BibitemShut
  {NoStop}%
\bibitem [{\citenamefont {Guerreiro}\ \emph {et~al.}(2014)\citenamefont
  {Guerreiro}, \citenamefont {Martin}, \citenamefont {Sanguinetti},
  \citenamefont {Pelc}, \citenamefont {Langrock}, \citenamefont {Fejer},
  \citenamefont {Gisin}, \citenamefont {Zbinden}, \citenamefont {Sangouard},\
  and\ \citenamefont {Thew}}]{guerreiro_nonlinear_2014}%
  \BibitemOpen
  \bibfield  {author} {\bibinfo {author} {\bibfnamefont {T.}~\bibnamefont
  {Guerreiro}}, \bibinfo {author} {\bibfnamefont {A.}~\bibnamefont {Martin}},
  \bibinfo {author} {\bibfnamefont {B.}~\bibnamefont {Sanguinetti}}, \bibinfo
  {author} {\bibfnamefont {J.}~\bibnamefont {Pelc}}, \bibinfo {author}
  {\bibfnamefont {C.}~\bibnamefont {Langrock}}, \bibinfo {author}
  {\bibfnamefont {M.}~\bibnamefont {Fejer}}, \bibinfo {author} {\bibfnamefont
  {N.}~\bibnamefont {Gisin}}, \bibinfo {author} {\bibfnamefont
  {H.}~\bibnamefont {Zbinden}}, \bibinfo {author} {\bibfnamefont
  {N.}~\bibnamefont {Sangouard}},\ and\ \bibinfo {author} {\bibfnamefont
  {R.}~\bibnamefont {Thew}},\ }\bibfield  {title} {\bibinfo {title} {Nonlinear
  {Interaction} between {Single} {Photons}},\ }\href
  {https://doi.org/10.1103/PhysRevLett.113.173601} {\bibfield  {journal}
  {\bibinfo  {journal} {Physical Review Letters}\ }\textbf {\bibinfo {volume}
  {113}},\ \bibinfo {pages} {173601} (\bibinfo {year} {2014})}\BibitemShut
  {NoStop}%
\bibitem [{\citenamefont {Akhmadaliev}\ \emph {et~al.}(2002)\citenamefont
  {Akhmadaliev}, \citenamefont {Kezerashvili}, \citenamefont {Klimenko},
  \citenamefont {Lee}, \citenamefont {Malyshev}, \citenamefont {Maslennikov},
  \citenamefont {Milov}, \citenamefont {Milstein}, \citenamefont {Muchnoi},
  \citenamefont {Naumenkov}, \citenamefont {Panin}, \citenamefont
  {Peleganchuk}, \citenamefont {Pospelov}, \citenamefont {Protopopov},
  \citenamefont {Romanov}, \citenamefont {Shamov}, \citenamefont {Shatilov},
  \citenamefont {Simonov}, \citenamefont {Strakhovenko},\ and\ \citenamefont
  {Tikhonov}}]{akhmadaliev_experimental_2002}%
  \BibitemOpen
  \bibfield  {author} {\bibinfo {author} {\bibfnamefont {S.~Z.}\ \bibnamefont
  {Akhmadaliev}}, \bibinfo {author} {\bibfnamefont {G.~Y.}\ \bibnamefont
  {Kezerashvili}}, \bibinfo {author} {\bibfnamefont {S.~G.}\ \bibnamefont
  {Klimenko}}, \bibinfo {author} {\bibfnamefont {R.~N.}\ \bibnamefont {Lee}},
  \bibinfo {author} {\bibfnamefont {V.~M.}\ \bibnamefont {Malyshev}}, \bibinfo
  {author} {\bibfnamefont {A.~L.}\ \bibnamefont {Maslennikov}}, \bibinfo
  {author} {\bibfnamefont {A.~M.}\ \bibnamefont {Milov}}, \bibinfo {author}
  {\bibfnamefont {A.~I.}\ \bibnamefont {Milstein}}, \bibinfo {author}
  {\bibfnamefont {N.~Y.}\ \bibnamefont {Muchnoi}}, \bibinfo {author}
  {\bibfnamefont {A.~I.}\ \bibnamefont {Naumenkov}}, \bibinfo {author}
  {\bibfnamefont {V.~S.}\ \bibnamefont {Panin}}, \bibinfo {author}
  {\bibfnamefont {S.~V.}\ \bibnamefont {Peleganchuk}}, \bibinfo {author}
  {\bibfnamefont {G.~E.}\ \bibnamefont {Pospelov}}, \bibinfo {author}
  {\bibfnamefont {I.~Y.}\ \bibnamefont {Protopopov}}, \bibinfo {author}
  {\bibfnamefont {L.~V.}\ \bibnamefont {Romanov}}, \bibinfo {author}
  {\bibfnamefont {A.~G.}\ \bibnamefont {Shamov}}, \bibinfo {author}
  {\bibfnamefont {D.~N.}\ \bibnamefont {Shatilov}}, \bibinfo {author}
  {\bibfnamefont {E.~A.}\ \bibnamefont {Simonov}}, \bibinfo {author}
  {\bibfnamefont {V.~M.}\ \bibnamefont {Strakhovenko}},\ and\ \bibinfo {author}
  {\bibfnamefont {Y.~A.}\ \bibnamefont {Tikhonov}},\ }\bibfield  {title}
  {\bibinfo {title} {Experimental {Investigation} of {High}-{Energy} {Photon}
  {Splitting} in {Atomic} {Fields}},\ }\href
  {https://doi.org/10.1103/PhysRevLett.89.061802} {\bibfield  {journal}
  {\bibinfo  {journal} {Physical Review Letters}\ }\textbf {\bibinfo {volume}
  {89}},\ \bibinfo {pages} {061802} (\bibinfo {year} {2002})}\BibitemShut
  {NoStop}%
\bibitem [{\citenamefont {Kolganov}(2023)}]{kolganov_real-time_2023}%
  \BibitemOpen
  \bibfield  {author} {\bibinfo {author} {\bibfnamefont {N.}~\bibnamefont
  {Kolganov}},\ }\bibfield  {title} {\bibinfo {title} {Real-time diagram
  technique for instantonic systems},\ }\href
  {https://doi.org/10.1007/JHEP10(2023)103} {\bibfield  {journal} {\bibinfo
  {journal} {Journal of High Energy Physics}\ }\textbf {\bibinfo {volume}
  {2023}},\ \bibinfo {pages} {103} (\bibinfo {year} {2023})}\BibitemShut
  {NoStop}%
\bibitem [{\citenamefont {Rajaraman}(1982)}]{rajaraman_solitons_1982}%
  \BibitemOpen
  \bibfield  {author} {\bibinfo {author} {\bibfnamefont {R.}~\bibnamefont
  {Rajaraman}},\ }\href@noop {} {\emph {\bibinfo {title} {Solitons and
  {Instantons}: {An} {Introduction} to {Solitons} and {Instantons} in {Quantum}
  {Field} {Theory}}}}\ (\bibinfo  {publisher} {North-Holland Publishing
  Company},\ \bibinfo {year} {1982})\BibitemShut {NoStop}%
\bibitem [{\citenamefont {Abramowitz}\ and\ \citenamefont
  {Stegun}(1965)}]{abramowitz_handbook_1965}%
  \BibitemOpen
  \bibfield  {author} {\bibinfo {author} {\bibfnamefont {M.}~\bibnamefont
  {Abramowitz}}\ and\ \bibinfo {author} {\bibfnamefont {I.~A.}\ \bibnamefont
  {Stegun}},\ }\href@noop {} {\emph {\bibinfo {title} {Handbook of
  {Mathematical} {Functions}: {With} {Formulas}, {Graphs}, and {Mathematical}
  {Tables}}}}\ (\bibinfo  {publisher} {Courier Corporation},\ \bibinfo {year}
  {1965})\BibitemShut {NoStop}%
\bibitem [{\citenamefont {Feynman}\ and\ \citenamefont
  {Vernon}(1963)}]{feynman_theory_1963}%
  \BibitemOpen
  \bibfield  {author} {\bibinfo {author} {\bibfnamefont {R.}~\bibnamefont
  {Feynman}}\ and\ \bibinfo {author} {\bibfnamefont {F.}~\bibnamefont
  {Vernon}},\ }\bibfield  {title} {\bibinfo {title} {The theory of a general
  quantum system interacting with a linear dissipative system},\ }\href
  {https://doi.org/https://doi.org/10.1016/0003-4916(63)90068-X} {\bibfield
  {journal} {\bibinfo  {journal} {Annals of Physics}\ }\textbf {\bibinfo
  {volume} {24}},\ \bibinfo {pages} {118} (\bibinfo {year} {1963})}\BibitemShut
  {NoStop}%
\bibitem [{\citenamefont {Gelmi}\ and\ \citenamefont
  {Jorquera}(2014)}]{gelmi_idsolver_2014}%
  \BibitemOpen
  \bibfield  {author} {\bibinfo {author} {\bibfnamefont {C.~A.}\ \bibnamefont
  {Gelmi}}\ and\ \bibinfo {author} {\bibfnamefont {H.}~\bibnamefont
  {Jorquera}},\ }\bibfield  {title} {\bibinfo {title} {{IDSOLVER}: {A} general
  purpose solver for $n$th-order integro-differential equations},\ }\href
  {https://doi.org/10.1016/j.cpc.2013.09.008} {\bibfield  {journal} {\bibinfo
  {journal} {Computer Physics Communications}\ }\textbf {\bibinfo {volume}
  {185}},\ \bibinfo {pages} {392} (\bibinfo {year} {2014})}\BibitemShut
  {NoStop}%
\bibitem [{\citenamefont {Silver}\ \emph {et~al.}(1990)\citenamefont {Silver},
  \citenamefont {Sivia},\ and\ \citenamefont
  {Gubernatis}}]{silver_maximum-entropy_1990}%
  \BibitemOpen
  \bibfield  {author} {\bibinfo {author} {\bibfnamefont {R.~N.}\ \bibnamefont
  {Silver}}, \bibinfo {author} {\bibfnamefont {D.~S.}\ \bibnamefont {Sivia}},\
  and\ \bibinfo {author} {\bibfnamefont {J.~E.}\ \bibnamefont {Gubernatis}},\
  }\bibfield  {title} {\bibinfo {title} {Maximum-entropy method for analytic
  continuation of quantum {Monte} {Carlo} data},\ }\href
  {https://doi.org/10.1103/PhysRevB.41.2380} {\bibfield  {journal} {\bibinfo
  {journal} {Physical Review B}\ }\textbf {\bibinfo {volume} {41}},\ \bibinfo
  {pages} {2380} (\bibinfo {year} {1990})}\BibitemShut {NoStop}%
\bibitem [{\citenamefont {Jarrell}\ and\ \citenamefont
  {Gubernatis}(1996)}]{jarrell_bayesian_1996}%
  \BibitemOpen
  \bibfield  {author} {\bibinfo {author} {\bibfnamefont {M.}~\bibnamefont
  {Jarrell}}\ and\ \bibinfo {author} {\bibfnamefont {J.~E.}\ \bibnamefont
  {Gubernatis}},\ }\bibfield  {title} {\bibinfo {title} {Bayesian inference and
  the analytic continuation of imaginary-time quantum {Monte} {Carlo} data},\
  }\href {https://doi.org/10.1016/0370-1573(95)00074-7} {\bibfield  {journal}
  {\bibinfo  {journal} {Physics Reports}\ }\textbf {\bibinfo {volume} {269}},\
  \bibinfo {pages} {133} (\bibinfo {year} {1996})}\BibitemShut {NoStop}%
\bibitem [{\citenamefont {Peskin}(2018)}]{peskin_introduction_2018}%
  \BibitemOpen
  \bibfield  {author} {\bibinfo {author} {\bibfnamefont {M.~E.}\ \bibnamefont
  {Peskin}},\ }\href@noop {} {\emph {\bibinfo {title} {An {Introduction} {To}
  {Quantum} {Field} {Theory}}}}\ (\bibinfo  {publisher} {CRC Press},\ \bibinfo
  {address} {Boca Raton},\ \bibinfo {year} {2018})\BibitemShut {NoStop}%
\bibitem [{\citenamefont {Vakhtel}\ and\ \citenamefont {van
  Heck}(2023)}]{vakhtel_quantum_2023}%
  \BibitemOpen
  \bibfield  {author} {\bibinfo {author} {\bibfnamefont {T.}~\bibnamefont
  {Vakhtel}}\ and\ \bibinfo {author} {\bibfnamefont {B.}~\bibnamefont {van
  Heck}},\ }\bibfield  {title} {\bibinfo {title} {Quantum phase slips in a
  resonant {Josephson} junction},\ }\href
  {https://doi.org/10.1103/PhysRevB.107.195405} {\bibfield  {journal} {\bibinfo
   {journal} {Physical Review B}\ }\textbf {\bibinfo {volume} {107}},\ \bibinfo
  {pages} {195405} (\bibinfo {year} {2023})}\BibitemShut {NoStop}%
\bibitem [{\citenamefont {Vakhtel}\ \emph {et~al.}(2024)\citenamefont
  {Vakhtel}, \citenamefont {Kurilovich}, \citenamefont {Pita-Vidal},
  \citenamefont {Bargerbos}, \citenamefont {Fatemi},\ and\ \citenamefont {van
  Heck}}]{vakhtel_tunneling_2024}%
  \BibitemOpen
  \bibfield  {author} {\bibinfo {author} {\bibfnamefont {T.}~\bibnamefont
  {Vakhtel}}, \bibinfo {author} {\bibfnamefont {P.~D.}\ \bibnamefont
  {Kurilovich}}, \bibinfo {author} {\bibfnamefont {M.}~\bibnamefont
  {Pita-Vidal}}, \bibinfo {author} {\bibfnamefont {A.}~\bibnamefont
  {Bargerbos}}, \bibinfo {author} {\bibfnamefont {V.}~\bibnamefont {Fatemi}},\
  and\ \bibinfo {author} {\bibfnamefont {B.}~\bibnamefont {van Heck}},\
  }\bibfield  {title} {\bibinfo {title} {Tunneling of fluxons via a {Josephson}
  resonant level},\ }\href {https://doi.org/10.1103/PhysRevB.110.045404}
  {\bibfield  {journal} {\bibinfo  {journal} {Physical Review B}\ }\textbf
  {\bibinfo {volume} {110}},\ \bibinfo {pages} {045404} (\bibinfo {year}
  {2024})}\BibitemShut {NoStop}%
\bibitem [{\citenamefont {Randeria}\ \emph {et~al.}(2024)\citenamefont
  {Randeria}, \citenamefont {Hazard}, \citenamefont {Di~Paolo}, \citenamefont
  {Azar}, \citenamefont {Hays}, \citenamefont {Ding}, \citenamefont {An},
  \citenamefont {Gingras}, \citenamefont {Niedzielski}, \citenamefont
  {Stickler}, \citenamefont {Grover}, \citenamefont {Yoder}, \citenamefont
  {Schwartz}, \citenamefont {Oliver},\ and\ \citenamefont
  {Serniak}}]{randeria_dephasing_2024}%
  \BibitemOpen
  \bibfield  {author} {\bibinfo {author} {\bibfnamefont {M.~T.}\ \bibnamefont
  {Randeria}}, \bibinfo {author} {\bibfnamefont {T.~M.}\ \bibnamefont
  {Hazard}}, \bibinfo {author} {\bibfnamefont {A.}~\bibnamefont {Di~Paolo}},
  \bibinfo {author} {\bibfnamefont {K.}~\bibnamefont {Azar}}, \bibinfo {author}
  {\bibfnamefont {M.}~\bibnamefont {Hays}}, \bibinfo {author} {\bibfnamefont
  {L.}~\bibnamefont {Ding}}, \bibinfo {author} {\bibfnamefont {J.}~\bibnamefont
  {An}}, \bibinfo {author} {\bibfnamefont {M.}~\bibnamefont {Gingras}},
  \bibinfo {author} {\bibfnamefont {B.~M.}\ \bibnamefont {Niedzielski}},
  \bibinfo {author} {\bibfnamefont {H.}~\bibnamefont {Stickler}}, \bibinfo
  {author} {\bibfnamefont {J.~A.}\ \bibnamefont {Grover}}, \bibinfo {author}
  {\bibfnamefont {J.~L.}\ \bibnamefont {Yoder}}, \bibinfo {author}
  {\bibfnamefont {M.~E.}\ \bibnamefont {Schwartz}}, \bibinfo {author}
  {\bibfnamefont {W.~D.}\ \bibnamefont {Oliver}},\ and\ \bibinfo {author}
  {\bibfnamefont {K.}~\bibnamefont {Serniak}},\ }\bibfield  {title} {\bibinfo
  {title} {Dephasing in {Fluxonium} {Qubits} from {Coherent} {Quantum} {Phase}
  {Slips}},\ }\href {https://doi.org/10.1103/PRXQuantum.5.030341} {\bibfield
  {journal} {\bibinfo  {journal} {PRX Quantum}\ }\textbf {\bibinfo {volume}
  {5}},\ \bibinfo {pages} {030341} (\bibinfo {year} {2024})}\BibitemShut
  {NoStop}%
\bibitem [{\citenamefont {Mencia}\ \emph {et~al.}(2024)\citenamefont {Mencia},
  \citenamefont {Lin}, \citenamefont {Cho}, \citenamefont {Vavilov},\ and\
  \citenamefont {Manucharyan}}]{mencia_integer_2024}%
  \BibitemOpen
  \bibfield  {author} {\bibinfo {author} {\bibfnamefont {R.~A.}\ \bibnamefont
  {Mencia}}, \bibinfo {author} {\bibfnamefont {W.-J.}\ \bibnamefont {Lin}},
  \bibinfo {author} {\bibfnamefont {H.}~\bibnamefont {Cho}}, \bibinfo {author}
  {\bibfnamefont {M.~G.}\ \bibnamefont {Vavilov}},\ and\ \bibinfo {author}
  {\bibfnamefont {V.~E.}\ \bibnamefont {Manucharyan}},\ }\bibfield  {title}
  {\bibinfo {title} {Integer fluxonium qubit},\ }\href
  {https://doi.org/10.1103/PRXQuantum.5.040318} {\bibfield  {journal} {\bibinfo
   {journal} {PRX Quantum}\ }\textbf {\bibinfo {volume} {5}},\ \bibinfo {pages}
  {040318} (\bibinfo {year} {2024})}\BibitemShut {NoStop}%
\bibitem [{\citenamefont {Burshtein}\ \emph {et~al.}(2025)\citenamefont
  {Burshtein}, \citenamefont {Shuliutsky}, \citenamefont {Kuzmin},
  \citenamefont {Manucharyan},\ and\ \citenamefont
  {Goldstein}}]{burshtein_zenodo16596445_2025}%
  \BibitemOpen
  \bibfield  {author} {\bibinfo {author} {\bibfnamefont {A.}~\bibnamefont
  {Burshtein}}, \bibinfo {author} {\bibfnamefont {D.}~\bibnamefont
  {Shuliutsky}}, \bibinfo {author} {\bibfnamefont {R.}~\bibnamefont {Kuzmin}},
  \bibinfo {author} {\bibfnamefont {V.}~\bibnamefont {Manucharyan}},\ and\
  \bibinfo {author} {\bibfnamefont {M.}~\bibnamefont {Goldstein}},\ }\bibfield
  {title} {\bibinfo {title} {arxiv:2410.23062: Data for fig. 5},\ }\href
  {https://doi.org/10.5281/zenodo.16596445} {10.5281/zenodo.16596445} (\bibinfo
  {year} {2025})\BibitemShut {NoStop}%
\bibitem [{\citenamefont {Gogolin}\ \emph {et~al.}(2004)\citenamefont
  {Gogolin}, \citenamefont {Nersesyan},\ and\ \citenamefont
  {Tsvelik}}]{gogolin_bosonization_2004}%
  \BibitemOpen
  \bibfield  {author} {\bibinfo {author} {\bibfnamefont {A.~O.}\ \bibnamefont
  {Gogolin}}, \bibinfo {author} {\bibfnamefont {A.~A.}\ \bibnamefont
  {Nersesyan}},\ and\ \bibinfo {author} {\bibfnamefont {A.~M.}\ \bibnamefont
  {Tsvelik}},\ }\href@noop {} {\emph {\bibinfo {title} {{Bosonization and
  Strongly Correlated Systems}}}}\ (\bibinfo  {publisher} {{Cambridge
  University Press, Cambridge}},\ \bibinfo {year} {2004})\BibitemShut {NoStop}%
\end{thebibliography}%
\end{document}